\documentclass[%
  prx,
  twocolumn,
  superscriptaddress,
  amsmath,
  amssymb,
  aps
]{revtex4-2}
\usepackage[colorlinks = True, linkcolor = blue, citecolor = blue]{hyperref}
\usepackage{graphicx}
\usepackage[normalem]{ulem}

\newcommand{\A}{\mathrm{A}}
\newcommand{\B}{\mathrm{B}}

\begin{document}
\title{Emergent Thermalization Thresholds in Unitary Dynamics of Inhomogeneously Disordered Systems}
\author{Soumya Kanti Pal}
\email{Equal contribution, Email: soumya.pal@tifr.res.in}
\affiliation{Department of Theoretical Physics, Tata Institute of Fundamental Research, Homi Bhabha Road, Mumbai 400005, India}
\author{C L Sriram}
\email{Equal contribution, Email: c.sriram@tifr.res.in}
\affiliation{Department of Theoretical Physics, Tata Institute of Fundamental Research, Homi Bhabha Road, Mumbai 400005, India}
\author{Shamik Gupta}
\email{shamik.gupta@theory.tifr.res.in}
\affiliation{Department of Theoretical Physics, Tata Institute of Fundamental Research, Homi Bhabha Road, Mumbai 400005, India}

\begin{abstract}
Inspired by the avalanche scenario for many-body localization (MBL) instability, we reverse the conventional set-up and ask whether a large weakly-disordered chain can thermalize a smaller, strongly-disordered chain when the composite system evolves unitarily. Using transport as a dynamical probe, we identify three distinct thermalization regimes as a function of the disorder strength of the smaller chain: (i) complete thermalization with self-averaging at weak disorder, (ii) realization-dependent thermalization with strong sample-to-sample fluctuations at intermediate disorder, and (iii) absence of thermalization at strong disorder. We find that for a fixed length of the smaller chain, the non-self-averaging regime broadens with the size of the weakly-disordered chain, revealing a nuanced interplay between disorder and system size. These results highlight how inhomogeneous disorder can induce emergent thermalization thresholds in closed quantum systems, providing direct access to disorder regimes where thermalization or its absence can be reliably observed.
\end{abstract}
\maketitle
\section{Introduction}
Thermalization is a fascinating phenomenon in nature, whereby isolated many-body systems evolving in time reach equilibrium. Specifically, the long-time averages of macroscopic observables become time-independent, settling at values characteristic of thermal equilibrium. Unraveling the origin of thermalization in isolated quantum systems remains a major pursuit in contemporary quantum many-body physics, particularly in light of compelling experimental evidence of thermalization in cold-atom systems~\cite{trotzky_nature,gring_science,schneider_nature,cheneau_nature,kinoshita,jurcevic_quasi_nature,langen_thermal_nature,langen_science,kaufman_science}. In this regard, a successful theoretical framework is the Eigenstate Thermalization Hypothesis (ETH)~\cite{ETH_Srednicki, Deutsch_ETH_original, ETH_review, Thermalization_Rigol,kurchan_eth,deutsch_eth}, which posits that mid-spectrum eigenstates of isolated, non-integrable quantum many-body Hamiltonians are thermal. ETH has been supported by extensive studies across a wide range of systems~\cite{Lea_rigol_chaos,Satos_Rigol_2,Thermalization_Rigol,Rigol_PRL_ETH1,Biroli_ETH,Gogolin,PhysRevE.89.042112,kim_testing,PhysRevX.14.031029,PhysRevLett.134.140404,PhysRevB.111.054303, Buosante_engineering_dynamics_with_disorder, Tomadin_localization_vs_disorder, TorresHerrera_dynamics_at_localization, Buijsman_energy_distribution_of_disordered_systems, Chavez_transport_with_long_range_hopping}. This raises a fundamental question: whether and under what conditions does the ETH tend to fail?  

\begin{figure}
    \centering
    \includegraphics[width=1\linewidth]{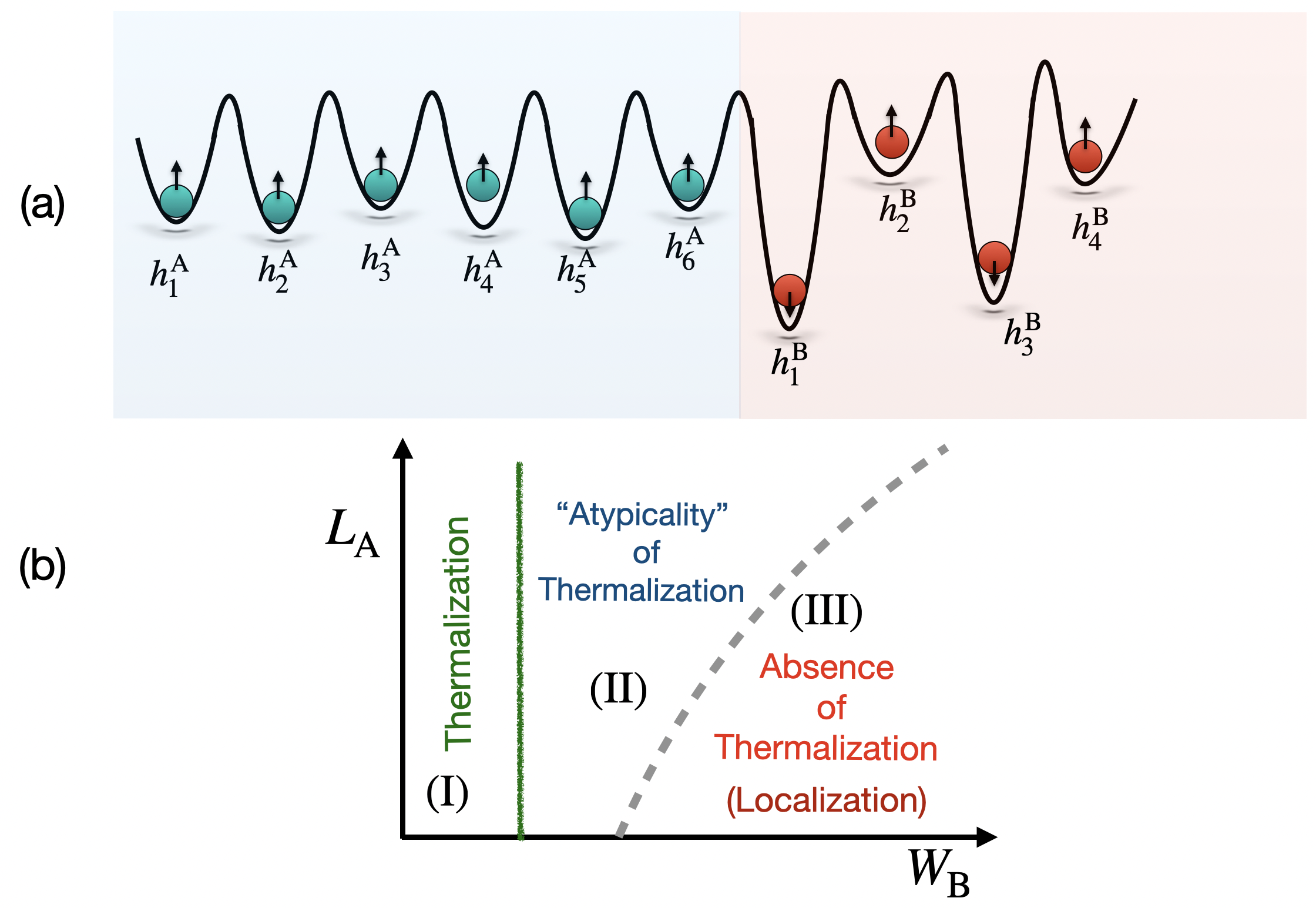}
    \caption{In panel~(a), we schematically show our set-up, where the blue region denotes the larger chain with weak disorder, while the red region represents the smaller, strongly disordered chain. The Hamiltonian of the system is given in Eq.~\eqref{eq:H}. Panel~(b) presents for a fixed value of $L_\B$, in the $(L_\A, W_\B)$ plane the three regimes (I), (II), and (III), detailed in the text, which show distinctively different thermalization behavior. The right boundary of regime~(II) shifts toward larger $W_\B$ values with increasing $L_\A$, indicating the growth of this regime with the size of the weakly-disordered chain.
   }
    \label{fig:setup}
\end{figure}

A well-established violation of ETH arises in isolated quantum many-body systems in the presence of quenched disorder: for sufficiently strong disorder, the system can enter a many-body localized (MBL) phase -- the interacting analogue of Anderson localization~\cite{Localization_Anderson} -- in which ETH no longer holds. While extensive numerical studies support the existence of MBL in finite-size systems \cite{MBL_Huse,PhysRevB.77.064426,Phenomenology_MBL_Nandkishore,MBL_Review_Nandkishore,Review_MBL_Abanin,PhysRevLett.117.040601,PhysRevLett.114.100601}, its stability in the limit of large systems remains an ongoing subject of investigation~\cite{Chaos_Challenges_Localization_Vidmar,Dynamical_Obstruction_MBL_Sels,Thermalization_of_Impurities_Polkovnikov}. One proposed mechanism for MBL instability is offered by the so-called avalanche scenario~\cite{PhysRevB.95.155129,David_luitz_PRL}. In strongly-disordered systems, small rare regions with anomalously-weak disorder -- ergodic grains -- can arise, which are characterized by dense spectra and fast internal dynamics, and which therefore can act as local thermal baths. When embedded in strongly-disordered regions and coupled to nearby localized sites, ergodic grains can induce resonant transitions in their surroundings, destabilizing the MBL phase and potentially triggering a cascade that thermalizes the entire system. This avalanche mechanism has been explored experimentally~\cite{PhysRevX.9.041014,leonard} as also in theory~\cite{avalanche_Thierry,David_luitz_PRL,Avalanche_MBL_Morningstar,PhysRevB.99.195145} with the ergodic grains modeled either in terms of Gaussian orthogonal ensemble (GOE) random matrix Hamiltonian~\cite{Deloc_in_MBL_DeRoeck,PhysRevB.95.155129,avalanche_Thierry} or as an infinite Markovian bath \cite{Bath_Induced_Delocalization_Dries, Avalanche_MBL_Morningstar} and the surrounding strongly-disordered regions treated as an open quantum system undergoing non-unitary evolution. However, recent work~\cite{PhysRevB.109.134202} suggests that a more complete description may require accounting for the full unitary evolution of the entire system. An interesting revelation reported in Ref.~\cite{David_luitz_PRL} was that a weakly-disordered chain of even 3 spins modeled as a GOE Hamiltonian could thermalize a strongly-disordered chain of 13 spins. Motivated by the aforesaid considerations, we now ask the reverse question: Can a large weakly-disordered chain $\A$ of length $L_\A$ and fixed disorder strength $W_\A$ always induce thermalization in a much smaller, strongly-disordered chain $\B$ of length $L_\B$ with disorder strength $W_\B$, under fully unitary evolution of the combined system? A schematic of our set-up is shown in Fig.~\ref{fig:setup}(a).

It is pertinent at this point to view our set-up vis-\`{a}-vis other set-ups explored in recent works. In \cite{PhysRevB.90.064203,PhysRevLett.114.117401}, the signatures of MBL were investigated in imperfect isolated settings, where the authors invoked the well-known ``l-bit" picture \cite{Phenomenology_MBL_Nandkishore,LIOM_Abanin,Ros2015}  in modelling the system, while considering the bath as an interacting phononic system. Using spectral statistics of local observables, a critical value of the coupling between the system and the bath was found above which MBL signature vanishes. In \cite{PhysRevB.92.014203}, the effect of a thermal bath coupled with an Anderson-localizing system was studied. By modelling the bath as a GOE Hamiltonian, it was revealed that while weak coupling facilitates transport, large-enough coupling induces an effect similar to localization, which was termed `Zeno localization.' In the same vein, there are other studies which address somewhat similar questions in the non-interacting Anderson-localization scenario~\cite{PhysRevB.105.L220203,PhysRevB.105.224208,PhysRevB.108.054201}. Here, for the interacting Heisenberg spin chains, we do not invoke the `l-bit' picture at strong disorder, and also do not model the thermal part as a GOE Hamiltonian. 

\begin{figure*}
    \begin{center}
        \includegraphics[width = 18.2 cm]{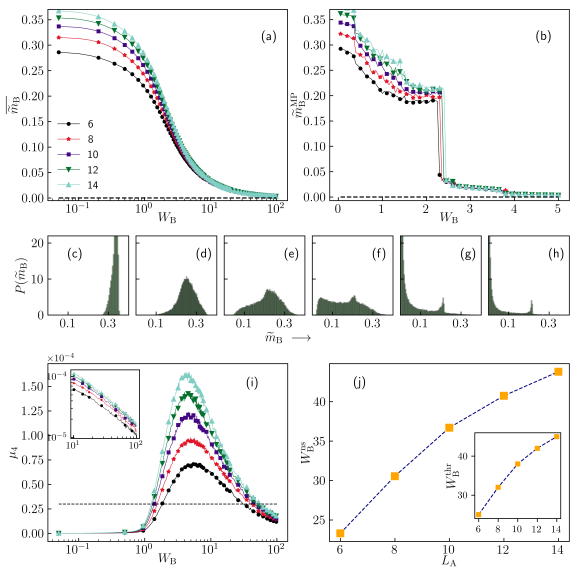}
        \caption{For the system~\eqref{eq:H}, (a) and (b) show respectively the disorder-average and the most probable value of the magnetization $\widetilde{m}_\B$ of system B of size $L_\B=4$, with system A having sizes $L_\A=6$ -- $14$ in steps of $2$. Panels (c)--(h) depict the probability distribution of $\widetilde{m}_\B$ for disorder strengths $W_\B=0.15, 1.05, 1.55, 2.3, 6.05, 12.2$, and $L_\A=10$. Panel (i) shows variation of $\mu_4\equiv \overline{(\widetilde{{m}}_\B-\overline{\widetilde{{m}}}_\B)^4}$ with $W_\B$ for various $L_\A$'s as in panel (a). We define $W_\mathrm{B}^\mathrm{ns}$ (indicated in panel (i) for $L_\A = 6$) as the range of $W_\B$ values over which $\mu_4$ takes up values above $0.3 \times 10^{-4}$ (horizontal dashed line in (i)). This is different from $W_\mathrm{B}^\mathrm{thr}$  (also shown in panel (i) for $L_\A = 6$), which is the value of $W_\mathrm{B}$ after which $\mu_4$ takes up values below $0.3 \times 10^{-4}$. Panel (j) shows $W_\B^{\mathrm{ns}}$ as a function of $L_\A$.  The inset in (i) shows the tail behavior, depicting distinct differences for different $L_\A$'s, while that in (j) shows as a function of $L_\A$ the value $W_\B^\mathrm{thr}$ of $W_\B$ beyond which one has inhibited spin transport from A to B. Every panel corresponds to $J=1$, $W_\A=0.1$ and $5\times 10^4$ disorder realizations.}
        \label{fig:fig_2}
    \end{center}
\end{figure*}

In this work, we reveal that the thermalization behavior of chain $\B$ is highly nuanced: for fixed $W_\A$, $L_\A$, and $L_\B$, we identify three distinct regimes of behavior depending on the value of $W_\B$, see Fig.~\ref{fig:setup}(b). (I) For small $W_\B$, thermalization occurs across all disorder realizations of chain $\B$, and macroscopic observables become self-averaging. (II) For intermediate values of $W_\B$, thermalization becomes realization-dependent, hence atypical: some realizations thermalize, while others do not. This results in pronounced sample-to-sample fluctuations and a breakdown of self-averaging, as illustrated in Fig.~\ref{fig:fig_2}(c)--(h), underscoring the limitations of disorder-averaged observables as reliable indicators of thermalization. (III) For large $W_\B$, absence of thermalization is observed across all disorder realizations. Remarkably, regime~(II) of broken self-averaging broadens with the size of chain~$\A$ while the length of the chain~$\B$ is kept fixed, as evidenced numerically in Fig.~\ref{fig:fig_2}(j), and qualitatively captured analytically using Fermi's Golden Rule analysis in Fig.~\ref{fig:Irwin_Hall}(b),(d). We clearly note that the proliferation of non-self-averaging in regime (II) is a finite-size effect that arises only in the limit $L_A \to \infty$ with fixed $L_B$; in the true thermodynamic limit with fixed $L_A/L_B$  and $L_A, L_B \to \infty$, regime (II) vanishes. Nonetheless, our analysis gives direct access to disorder regimes for observing thermalization or absence thereof in a typical finite-size sample, and warns against probing of thermalization in regimes in which strong sample-to-sample variations do not admit an unequivocal statement on thermalization based on naive single-sample measurements. 

Quenched disorder results in a range of intriguing phenomena in classical and quantum systems, e.g.,   spin-glass phases~\cite{Parisi-Mezard,spin-glass-quantum}, aging~\cite{Leticia-Kurchan-aging}, and Anderson localization~\cite{Localization_Anderson}. An emergent hallmark in such systems is the absence of self-averaging, wherein macroscopic observables retain a dependence on microscopic disorder realizations even in the thermodynamic limit. This phenomenon has been extensively studied in classical equilibrium systems~\cite{Self_Averaging_Domany,Self_Averaging_Aharony}, particularly near critical points, where strong sample-to-sample fluctuations are known to arise. Even in nonequilibrium settings, steady-state transitions \cite{Derrida} and transport properties \cite{TASEP_NSA,Transport_NSA} may exhibit strong dependence on individual disorder realizations. Recent studies in quantum many-body systems~\cite{PhysRevB.102.094310,PhysRevB.110.075138,bouverot2024random} have highlighted the absence of self-averaging in indicators such as inverse participation ratio, correlation functions, survival probabilities, in disordered XXZ chain and GOE random matrices. These studies focus on homogeneously-disordered systems, in contrast to our system that has inhomogeneous disorder. Within our set-up, we demonstrate that non-self-averaging behavior can also emerge in a distinctively different class of indicators than the ones above, highlighting the subtle role of disorder in thermalization under fully unitary evolution.

The paper is organized as follows. In Section~\ref{sec:model}, we introduce our model comprising two disordered Heisenberg spin chains coupled via a boundary spin, and describe the set-up of our numerical experiment that probes spin transport between the chains to investigate the fate of thermalization in the smaller chain. Section~\ref{sec:numerics} presents the numerical results, where we identify and characterize the three distinct thermalization regimes (I), (II), and (III) discussed above. In Section~\ref{sec:analytics}, we focus on regime (II): we first study its growth with increasing $L_\A$ based on our numerical findings. We then qualitatively explain this behavior using Fermi’s Golden Rule and extreme value theory, which further allows us to estimate the threshold disorder strength $W_\B^{\mathrm{thr}}$ separating regimes (II) and (III) for given values of $L_\A$ and $L_\B$. We conclude in Section~\ref{sec:conclusion} with a summary of our results, along with remarks on possible extensions and experimental realizations of the proposed set-up.

\section{Model and numerical probe of thermalization}\label{sec:model}
To demonstrate our results summarized above, we consider a composite quantum system comprising two interacting many-body spin-$1/2$ Heisenberg chains with disordered onsite fields: a larger weakly-disordered chain $\A$ and a smaller chain $\B$ with tunable disorder strength, see Fig.~\ref{fig:setup}(a). The two chains are coupled via a shared boundary spin. We also examine the case where the larger chain is clean. To diagnose thermalization, we design a numerical experiment that probes the transport of $z$-magnetization from chain $\A$ to chain $\B$. To proceed, consider an open Heisenberg chain of length $L_\mathrm{B}$ with disordered onsite fields $\{h_i^\mathrm{B}\}$, which is in contact with another open Heisenberg chain of length $L_\mathrm{A}$ with disordered onsite fields $\{h_i^\mathrm{A}\}$, with $L_\B \ll L_\A$. The Hamiltonian is  
\begin{align}
    \hat{H} = J\sum_{i =1}^{L-1} \hat{\Vec{S}}_i \cdot \hat{\Vec{S}}_{i+1}  + \sum_{i \in \A} h_i^\A
    \hat{S}^z_i + \sum_{j \in \B} h_j^\B \hat{S}^z_j,
    \label{eq:H}
\end{align}  
where the coupling strength \( J > 0 \) is uniform throughout, both within and across the two chains. The total system size is \( L \equiv L_\mathrm{A} + L_\mathrm{B} \). The on-site disorder fields \( h_i^\A \) and \( h_j^\B \) are independently drawn from uniform distributions over \( [-W_\A, W_\A] \) and \( [-W_\B, W_\B] \), respectively, with \( W_\A, W_\B > 0 \). We fix \( W_\A = 0.1J \) to represent weak disorder in chain A, while \( W_\B < J \) and \( W_\B \gg J \) correspond to weak and strong disorder in chain B, respectively, with intermediate values lying in between.
The Hamiltonian~\eqref{eq:H} has global $\hat{S}^z$ symmetry: $[\hat{H}, \hat{S}^z] = 0$, with $\hat{S}^z \equiv (1/L)\sum_{i=1}^{L} \hat{S}_i^z$. Let us define individual magnetizations $\hat{m}_\A \equiv (1/L_\A) \sum_{i \in \A} \hat{S}^z_i$ and $\hat{m}_\B \equiv (1/L_\B) \sum_{i \in \B} \hat{S}^z_i$, which are non-conserved quantities evolving in time with the constraint
\begin{align}
L_\A m_\A(0) + L_\B m_\B(0) = L_\A m_\A(t) + L_\B m_\B(t), 
\label{eq:m-global-constraint}
\end{align}
with $m_\A(t) \equiv \langle \psi(t)| \hat{m}_\A\otimes \hat{I}_\B| \psi(t) \rangle,~m_\B(t) \equiv \langle \psi(t)| \hat{I}_\A \otimes \hat{m}_\B| \psi(t) \rangle$, $\hat{I}_\A$ and $\hat{I}_\B$ being identity operators, and $|\psi(t)\rangle=e^{-i\hat{H}t}|\psi_0\rangle$ the state of A+B system at time $t$ with initial state $|\psi_0\rangle$~\footnote{We work in units in which the reduced Planck's constant is unity.}. We choose $|\psi_0\rangle$ as an eigenstate of $\hat{S}^z$, with $m_\A(0)\neq 0$, $m_\B(0)=0$ (chain A is initially magnetized, while chain B is not). We choose the representative case $m_\A(0) = 1$. For a given disorder realization, a quantity of interest is the long-time average of $m_\B(t)$:
\begin{align}
    \widetilde{{m}}_\B \equiv \lim_{t \to \infty} \frac{1}{t} \int_{0}^{t} m_\B(t') ~\mathrm{d}t'.
\end{align}
Expanding $|\psi(t)\rangle$ in energy eigenbasis, $|\psi(t)\rangle=\sum_n c_n e^{-iE_nt}|n\rangle$, it is clear that $\widetilde{{m}}_\B$ is given by the diagonal ensemble: $\widetilde{{m}}_\B=\sum_n |c_n|^2 \langle n|\hat{I}_\A\otimes \hat{m}_B|n\rangle$. Clearly, $\widetilde{{m}}_\B$ is a random variable varying from one disorder realization to another, and it is natural to propose its disorder-average $\overline{\widetilde{{m}}}_\B$~\footnote{Averaging over disorder in $\B$ (for fixed disorder realization in $\A$), then over disorder realizations in $\A$, is equivalent to averaging over disorder in the full chain, since $\A$ and $\B$ have independent disorder.} becoming non-zero as a signature of transport from A to B for our choice of $|\psi_0\rangle$~\footnote{The corresponding criterion for $m_\B(0) \ne 0$ will be to have $\overline{|m_\B(0) - \protect\widetilde{m}_\B|} \ne 0$.}. In case of inhibited transport, evidently there is no hope for chain $\B$ to thermalize. 

\section{Numerical results}\label{sec:numerics}
We now present our numerical results on $\overline{\widetilde{{m}}}_\B$ as a function of disorder strength $W_\B$ and for fixed $L_\B$, Fig.~\ref{fig:fig_2}; here, panel (a) shows that at low disorder, $\overline{\widetilde{{m}}}_\B$ takes non-zero values, indicating transport from A to B and hence thermalization of chain B; these values increase with increasing $L_\A$, owing to constraint~\eqref{eq:m-global-constraint}. With increase of $W_\B$, $\overline{\widetilde{{m}}}_\B$ asymptotically approaches zero, signaling inhibition of transport and hence absence of thermalization in chain $\B$ at large-enough $W_\B$. Cautioned by the fact that classical disordered systems are notorious for showing strong atypical behaviour~\cite{Self_Averaging_Aharony, Self_Averaging_Domany}, we investigate the typicality of transport in our system by comparing the average $\overline{\widetilde{{m}}}_\B$ \textit{vis-\`{a}-vis} the most probable value $\widetilde{m}_\B^\mathrm{MP}$. Figure~\ref{fig:fig_2}(b) shows a strikingly different behavior. At low disorder, one has clear signatures of spin transport. At high disorder, one has inhibited transport. However, the threshold $W_\B$ that inhibits transport is an order-of-magnitude different than the one implied by panel (a). This latter fact warrants a detailed analysis of the probability distribution $P(\widetilde{{m}}_\B)$ as one tunes $W_\B$. 

The results for $P(\widetilde{{m}}_\B)$, Fig.~\ref{fig:fig_2}(c)--(h), depict a dramatic change in its form, from ones at low and high disorder that are narrow and have a well-defined dominant peak, to those at intermediate disorder that are broad and without a single well-defined peak. The latter fact explains the absence of observed typicality ($\widetilde{m}_\B^\mathrm{MP}$ behaving differently from $\overline{\widetilde{{m}}}_\B$), and implies that $\widetilde{{m}}_\B$ is strongly non-self-averaging at intermediate disorder. This last feature is intriguing, given that the underlying disorder distribution does not have heavy tails, and, consequently, the disordered fields are self-averaging. To demonstrate this point, we have checked that our results hold also when the field values are sampled from a Gaussian distribution, see Appendix~\ref{APP-Gaussian}. The non-self-averaging feature of $\widetilde{{m}}_\B$ is characterized in terms of the fourth central moment $\mu_4 \equiv \overline{(\widetilde{{m}}_\B-\overline{\widetilde{{m}}}_\B)^4}$, which captures the broadness of $P(\widetilde{{m}}_\B)$ more sensitively than the variance; the higher the value of $\mu_4$, more broad is $P(\widetilde{{m}}_\B)$ and more non-self-averaging is $\widetilde{{m}}_\B$. The behavior of $\mu_4$ versus $W_\B$ with fixed $L_\B$ and for various $L_\A$'s, shown in Fig.~\ref{fig:fig_2}(i), exhibits enhanced $\mu_4$-values above zero for a wide range of $W_\B$.  Combined with panels (a) and (b), panel (i) implies: (1) Spin transport and thermalization at weak disorder, (2) Transport and thermalization becoming realization-dependent at intermediate disorder, and (3) Inhibited transport and no thermalization at strong disorder. Self-averaging holds in regimes (1) and (3), but not in (2). The most striking aspect of the behavior of $\mu_4$ is the pronounced growth with increasing $L_\A$ of the range of $W_\B$ for which behavior (2) holds. Estimating this range as the one over which $\mu_4$ takes values above a non-zero threshold and plotting it versus $L_\A$, one obtains Fig.~\ref{fig:fig_2}(j), which conveys a surprising result: the range of $W_\B$ for which chain B is non-self-averaging increases monotonically with the length of the chain A in contact. The existence of these regions are schematically demonstrated in Fig.~\ref{fig:setup}(b)~\footnote{We note that our results remain qualitatively unchanged for the ring geometry, where the two chains are coupled at two contact points.}. Our results also hold when system $\A$ has a uniform field~$h^\A$, as demonstrated in Appendix~\ref{App-uniform-field}. We note that the schematic phase diagram presented for our setup differs significantly from the case with $L_A/L_B$ fixed and $L_A, L_B \to \infty$. In this latter limit, region~(II) vanishes; however, for the system sizes accessible to us numerically, we cannot conclusively rule out the existence of region~(III). Importantly, in realistic scenarios involving finite-size systems, the natural outcome is the one shown in Fig.~\ref{fig:setup}(a), where our analysis uncovers three distinct regimes of thermalization, including a proliferated zone of non-self-averaging.

\section{When transport becomes typical: A perturbative approach}\label{sec:analytics}
A relevant question is: What is the threshold value of $W_\B$ for inhibited spin transport from A to B? In view of $\widetilde{m}_{\B}$ being non-self-averaging, we modify the hitherto-mentioned criterion for inhibited transport, namely, $\overline{\widetilde{m}}_{\B}=0 $, to $\widetilde{m}_{\B}=0$ for almost all realizations, which ensures typicality of inhibition. In $W_\B \gg J$ regime, where inhibition is typical, Fig.~\ref{fig:fig_2}(a),(b), this modified criterion is met beyond a threshold $W_\B$ for which $\mu_4 =0$; In practical terms, one stipulates that $\mu_4$ takes up a small value of order zero and considers the corresponding $W_\B$ as the threshold beyond which inhibited transport is guaranteed for almost all realizations. As in the inset of Fig.~\ref{fig:fig_2}(j), the threshold $W_\B^\mathrm{thr}$ increases with $L_\A$, indicating that weakly-disordered systems of larger sizes are conducive to transport from A to B. In this section, we present a perturbative approach based on the Fermi's Golden Rule to estimate the threshold $\tilde{W}_\B^\mathrm{thr}$. 

\subsection{Simplest case: $L_{\B} =1$ }
We now study analytically system~\eqref{eq:H} by using perturbation theory in view of $L_\A \gg L_\B$. Based on our observations for inhibited transport, we first consider $L_\B=1, W_\B \gg J$, and, to make the analysis general, we take $W_\A < J$. The model is rewritten as 
\begin{align}
    \hat{\widetilde{H}} = J\sum_{i =1}^{L_\A-1} \hat{\Vec{S_i}} \cdot \hat{\Vec{S}}_{i+1}  +  \sum_{i=1}^{L_\A} h_i^\A \hat{S_i^z} + J \hat{\Vec{S_0}} \cdot \hat{\Vec{S}}_1+ W_\B \hat{S_0^z},
    \label{eq:H_P}
\end{align} 
denoting the single site of system B by index $0$ with onsite field $W_\B$. In the interaction picture with respect to $W_\B\hat{S}_0^z$, the interaction Hamiltonian $\hat{\widetilde{H}}_{I}=e^{iW_\B \hat{S_0^z}t}(\hat{\widetilde{H}} - W_\B \hat{S}_0^z) e^{-iW_\B \hat{S_0^z}t}$ has a Floquet form:
\begin{align}{\label{eq:H_mod}}
    \hat{\widetilde{H}}_{I} &= \hat{H}_{L_\A} +\hat{H}^\prime;~\hat{H}^\prime=\frac{ J}{2} (e^{-i W_\B t} \hat{S}_1^+ \hat{S}_0^- + e^{i W_\B t} \hat{S}_1^- \hat{S}_0^+);\nonumber \\\\
    \hat{H}_{L_\A} &= J\sum_{i=1}^{L_\A-1} \hat{\vec{S}}_i \cdot \hat{\vec{S}}_{i+1} + \sum_{i=1}^{L_\A} h_i^\A \hat{S}_i^z+J\hat{S}_0^z \hat{S}_1^z. \nonumber
\end{align}
With $L_\A\gg 1$, $\hat{H}^\prime$ acts as a weak perturbation to $\hat{H}_{L_\A}$, due to a time-scale separation between the dynamics generated by the two terms, guaranteed from the behavior of the ratio of their norms: $||\hat{H}^\prime||/||\hat{H}_{L_\A}|| \sim 1/{L_\A}$, with the norm defined as  $|| \hat{H} ||\equiv \max\{Sp(\hat{H})\}$, i.e., the largest eigenvalue of the operator spectrum $Sp(\hat{H})$. 

Starting with our choice of the initial state, dynamics generated by $\hat{H}_{L_\A}$ conserves $m_\A$ and $m_\B$ as a consequence of $[\hat{H}_{L_\A}, \hat{S}^z]=0$, while it is the perturbation $\hat{H}^\prime$ that leads to a change in $m_\B$, and hence, in $m_\A$ owing to constraint~\eqref{eq:m-global-constraint}; this implies spin transport from A to B. Note that $[\hat{H}^\prime, \hat{S}^z]=0$ constrains transitions to take place between different eigenstates of $\hat{H}_{L_\A}$ within the same $S^z$ sector. Moreover, any given $S^z$ sector is not confined within specific energy ranges, but is spread across the full spectrum $Sp(\hat{H}_{L_\A})$. Spin transport due to $\hat{H}^\prime$ is quantified by studying the transition it induces between eigenstates of $\hat{H}_{L_\A}$. Our next objective is to obtain conditions for the corresponding rate of transition to become zero, which in turn implies inhibition of transport. We now compute this rate by invoking first-order time-dependent perturbation theory~\cite{sakurai}, and estimate the sufficient value of $W_\B$ beyond which the rate is zero. The corresponding rate, given by the Fermi's Golden Rule (FGR)  for transition from eigenstate $|\alpha \rangle$ to any other eigenstate $|\beta \rangle$ of $\hat{H}_{L_\A}$, reads as~(see Appendix~\ref{App-FGR-derivation})
\begin{align}
    \Gamma_{\alpha\to [\beta]} &\propto \sum_\beta \Biggl \{|A_{\beta \alpha}|^2 \delta_+ + |A_{\beta\alpha}'|^2 \delta_- \Biggr \},
    \label{eq:Single_impurity_FGR}
\end{align}
with $\delta_\pm \equiv \delta(\omega_{\beta \alpha} \pm W_\B)$, $|A_{\beta \alpha}| = \langle \beta|\hat{S}_1^- \hat{S}_0^+|\alpha \rangle$, $|A'_{\beta \alpha}| = \langle \beta|\hat{S}_1^+ \hat{S}_0^-|\alpha \rangle$, $\omega_{\beta \alpha} \equiv E_\beta - E_\alpha$ with $E_\beta, E_\alpha$ eigenvalues of $\hat{H}_{L_\A}$ for eigenstates $|\beta \rangle$ and $|\alpha \rangle$, respectively. The FGR-approach, describing decay of an initial state into a continuum of final states, may not apply when one has a finite system and the final states do not form a continuum~\cite{PhysRevLett.129.140402}. Our analysis with $L_\A \gg L_\B$ creates a quasi-continuum for initial states in B to decay, justifying the applicability of the FGR-approach.

It follows from Eq.~(\ref{eq:Single_impurity_FGR}) that $\Gamma$ is zero provided $\delta(\omega_{\beta \alpha} \pm W_\B)=0$, the condition for which is $W_\B > \Delta$, with $\Delta \equiv \max\{Sp(\hat{H}_{L_\A})\} - \min\{Sp(\hat{H}_{L_\A})\}$~\footnote{The condition $W_\B < \mathrm{min}(\omega_{\beta \alpha})$ will also ensure that $\Gamma=0$; however, it turns out that $\omega_{\beta \alpha}$ is of
order $\sqrt{L_\A}2^{-L_\A}$~\cite{MBL_Huse}, which vanishes in the limit $L_\A \to \infty$. Then, in this limit, which is also what we will be interested in, the condition $W_\B < \mathrm{min}(\omega_{\beta \alpha})$ will never be met in view of the fact that we have $W_\B>0$.}. Transitions $\beta \leftrightarrow \alpha$ take place within the same $S^z$ sector. Nevertheless, $S^z$ sectors being spread across $Sp(\hat{H}_{L_\A})$ implies that one has to consider the full spectrum in stipulating the condition for inhibited transport. Noting that $\hat{H}_{L_\A}$ involves the random fields $\{h_i^\A\}$ uniform in $[-W_\A,W_\A]$, we ask: for given $W_\B$, what is the probability $P\equiv P(W_\B>\Delta)$ to have a disorder realization $\{h_i^\A\}$ that generates an $\hat{H}_L$ with spectral gap satisfying $W_\B>\Delta$? This probability being non-zero implies $\Gamma=0$, i.e., inhibited transport for given $W_\B$ and $L_\A$. For a given realization $\{h_i^\A\}$, while an explicit expression for $\Delta$ is not known, the problem is circumvented by putting an upper bound on $\Delta$.

To proceed, we write the unperturbed Hamiltonian as $\hat{H}_{L_\A}=\hat{H}_0+\hat{H}_1$, with 
\begin{align}
    \hat{H}_0 = J\sum_{i=1}^{L_\A-1} \hat{\vec{S}}_i \cdot \hat{\vec{S}}_{i+1} + J\hat{S}_0^z \hat{S}_1^z;~~\hat{H}_1=\sum_{i=1}^{L_\A} h_i^\A \hat{S}_i^z,
    \label{eq:H_L}
\end{align}
and denote the corresponding spectra as $\{\mu_\alpha\}\equiv Sp(\hat{H}_{L_\A})$, $\{\nu_\alpha\} \equiv Sp(\hat{H}_0)$ and $\{\lambda_\alpha \} \equiv Sp(\hat{H}_1)$, where we have $\alpha= 1, \ldots, 2^{L_\A+1}$.  For $Sp(\hat{H}_1)$, the maximum and minimum eigenvalues are $\lambda \equiv \max_{\alpha} \{ \lambda_\alpha \} = \sum_{i=1}^{L_\A} |h_i^\A|$  and $\tilde{\lambda} \equiv \min_{\alpha} \{ \lambda_\alpha \}= -\sum_{i=1}^{L_\A} |h_i^\A|$. Similarly, for $Sp(\hat{H}_0)$, $\nu \equiv \max_\alpha \{\nu_\alpha\} = JL_\A/4$, which corresponds to the maximum-energy configuration with all spins pointing along $z$~\cite{tasaki}, while $\tilde{\nu} \equiv \min_{\alpha} \{\nu_\alpha \}=-JL_\A(1/4+ 1/\pi)$, valid in the large $L_\A$-limit~\cite{AnnPhys_1961_Leib_Shultz_Mattis}. We invoke the Weyl's inequality: the largest and the smallest eigenvalue of sum $C$ of two Hermitian matrices $A$ and $B$ satisfy $\lambda_C \le \lambda_A+\lambda_B$ and $\tilde{\lambda}_C \ge \tilde{\lambda}_A+\tilde{\lambda}_B$~ \cite{Inequality_Weyl, Weyl_Inequality_Tao}. We thus obtain the bounds $\mu\equiv \max_\alpha\{\mu_\alpha\} \leq JL_\A/4 + (1/2)\sum_{i=1}^{L_\A} |h_i^\A| $ and $\tilde{\mu}\equiv \min_\alpha\{\mu_\alpha\} \geq -JL_\A(1/4 + 1/\pi) - (1/2)\sum_{i=1}^{L_\A} |h_i^\A|$. This readily yields an upper bound
\begin{align} \label{eq:upper bound}
    \Delta \leq \sum_{i=1}^{L_\A} (|h_i^\A| + J/2 + J/\pi) \equiv \widetilde{\Delta}.
\end{align}

Following our earlier line of argument, we now use $\widetilde{\Delta}$ in Eq.~\eqref{eq:upper bound} as a reasonable substitute for $\Delta$ in determining $P$ as $P=P(W_\B>\widetilde{\Delta})$ for given $W_\B$ and various $L_\A$'s. Evidently, $\widetilde{\Delta}$ follows a generalized Irwin-Hall distribution \cite{Irwin_Hall}, and $P(W_\B>\widetilde{\Delta})$ is the cumulative distribution function that can be computed as~(see Appendix~\ref{APP-Irwin-Hall}) 
\begin{align}
    P=  \sum_{k=0}^{L_\A}  \frac{(-1)^k  {L_\A \choose k}}{{W_\A}^{L_\A} L_\A!}\biggl[W_\B - W_\A k - \left(\frac{J}{2} + \frac{J}{\pi}\right)L_\A \biggr]^{L_A}_+,
    \label{eq:P-expression}
\end{align}
where $x_+=x$ if $x \geq 0$ and is zero otherwise. This expression holds for $W_B \in [(J/2 + J/\pi)L_\A, (J/2 + J/\pi + W_\A)L_\A]$, while $P = 0$ for $W_\B < (J/2 + J/\pi)L_\A$ and $P=1$ for $W_\B > (J/2 + J/\pi + W_\A)L_\A$. Figure~\ref{fig:Irwin_Hall}(a) shows a plot of $P$ versus $W_\B$ for various $L_\A$'s, and establishes that the value of $W_\B$ ensuring $P>0$ increases with increasing $L_\A$. It follows that to have $\Gamma=0$, i.e., inhibition of transport, one has to tune $W_\B$ to higher values for larger $L_\A$'s, as evident from Fig.~\ref{fig:Irwin_Hall}(b), where the threshold $\tilde{W}_\B^\mathrm{thr}$ above which $P>0$ shows a monotonic growth with $L_\A$. 

\subsubsection{Behaviour of the matrix elements}
\begin{figure}[!h]
    \begin{center}
        \includegraphics[width = 0.4\textwidth]{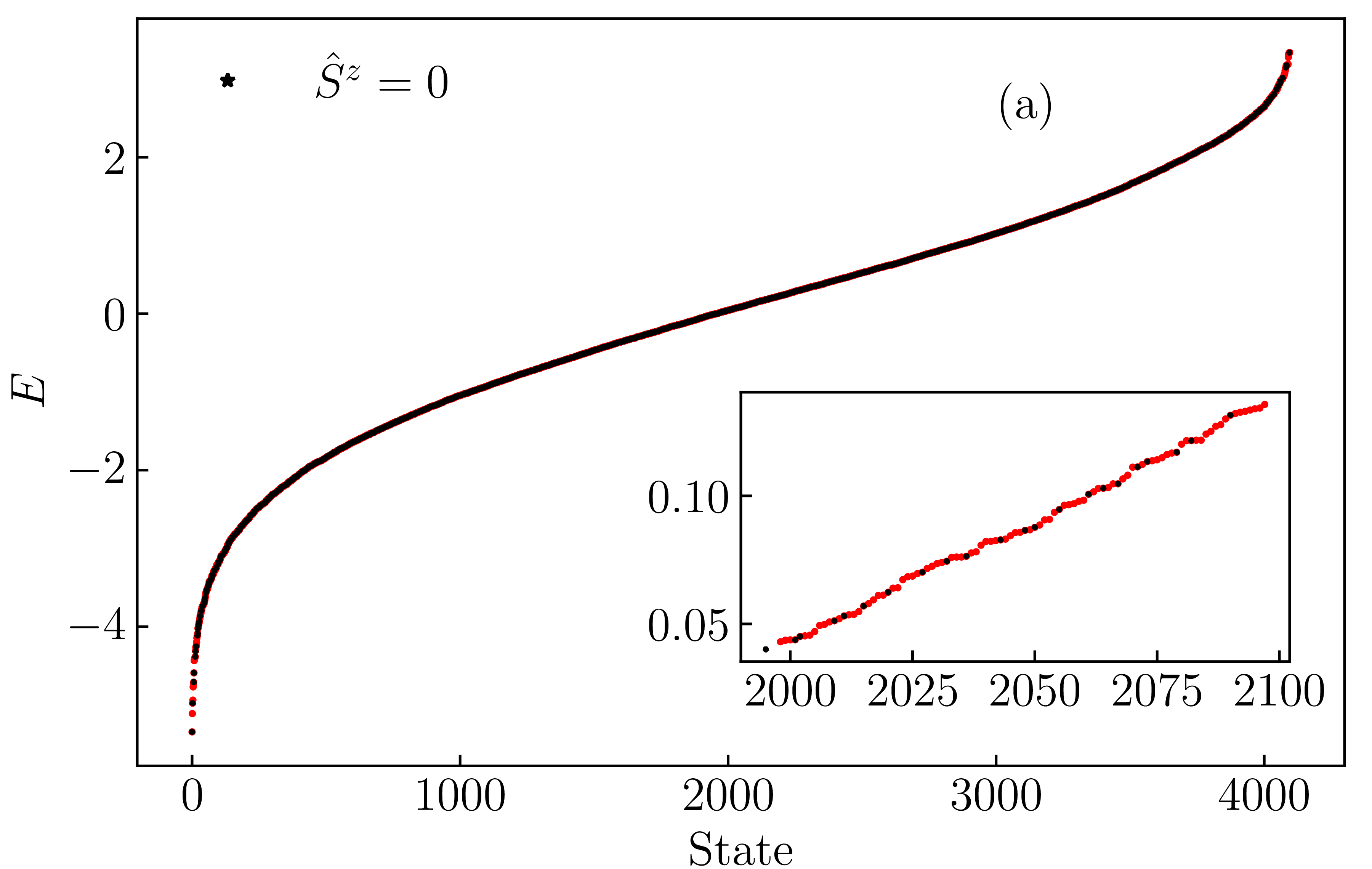}
        \includegraphics[width = 0.5\textwidth]{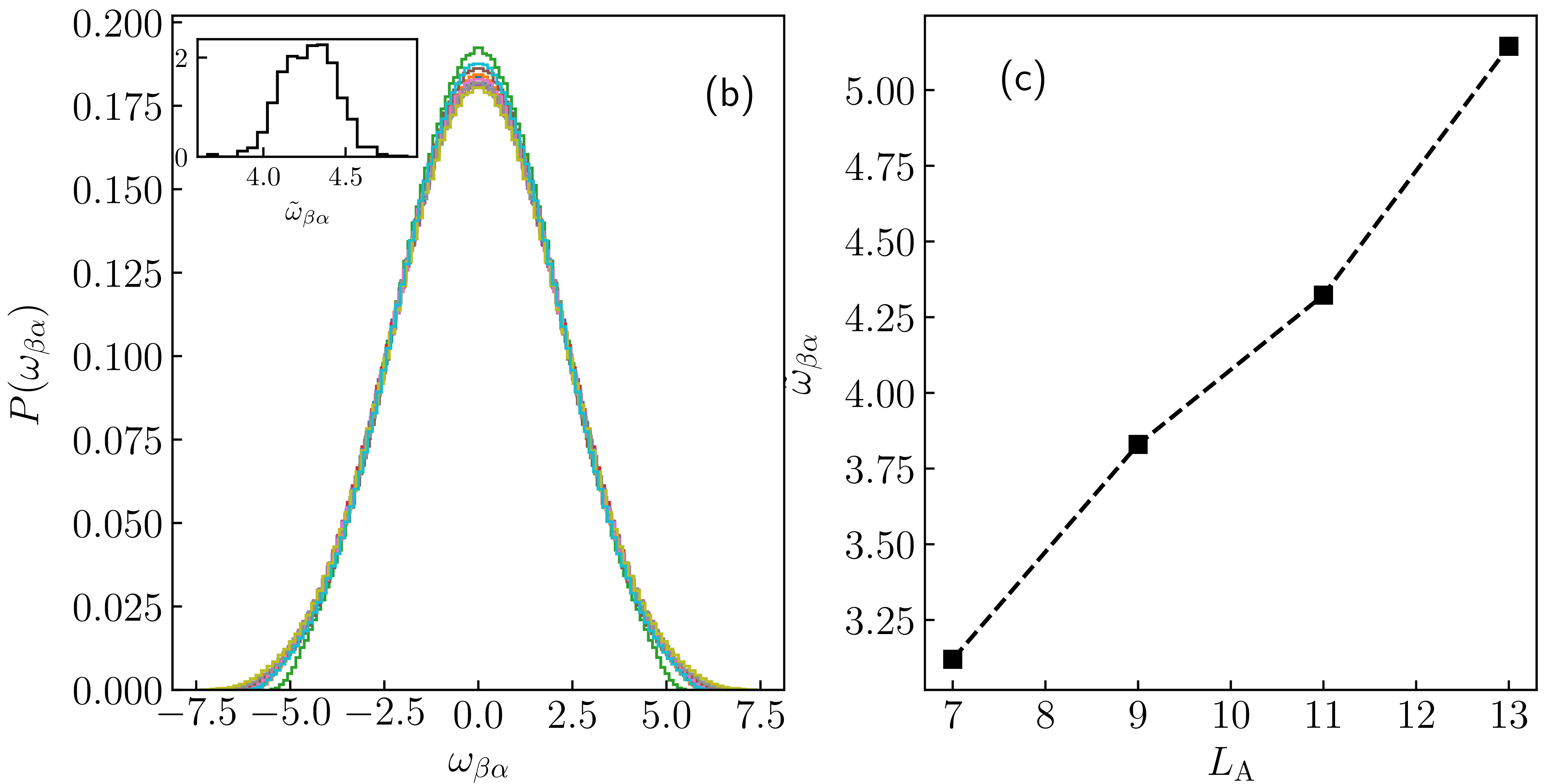}
        \caption{For $L_\A = 11$ and $W_\A = 0.5$, panel (a) shows the eigenstates of $H_{L_\A}$ (red) with the states belonging to the $S^z = 0$ sector (black). Inset is a blow up of the middle part of the spectrum; Panel (b) on the other hand shows the distribution $P(\omega_{\beta \alpha})$ for 10 disorder realizations $\{ h_i^\A \}$. The inset shows the distribution of $\tilde{\omega}_{\beta \alpha}\equiv \overline{\omega_{\beta \alpha}^2} - (\overline{\omega_{\beta \alpha}})^2$ calculated for $1000$ realizations of $\{h_i^\A \}$. Panel (c) shows $\tilde{\omega}_{\beta \alpha}$ as a function of $L_\A$, with $L_\B=1$. We have set $J=1$.}
        \label{fig:App_4}
    \end{center}
\end{figure}

\begin{figure}
    \begin{center}
        \includegraphics[width=8.6cm]{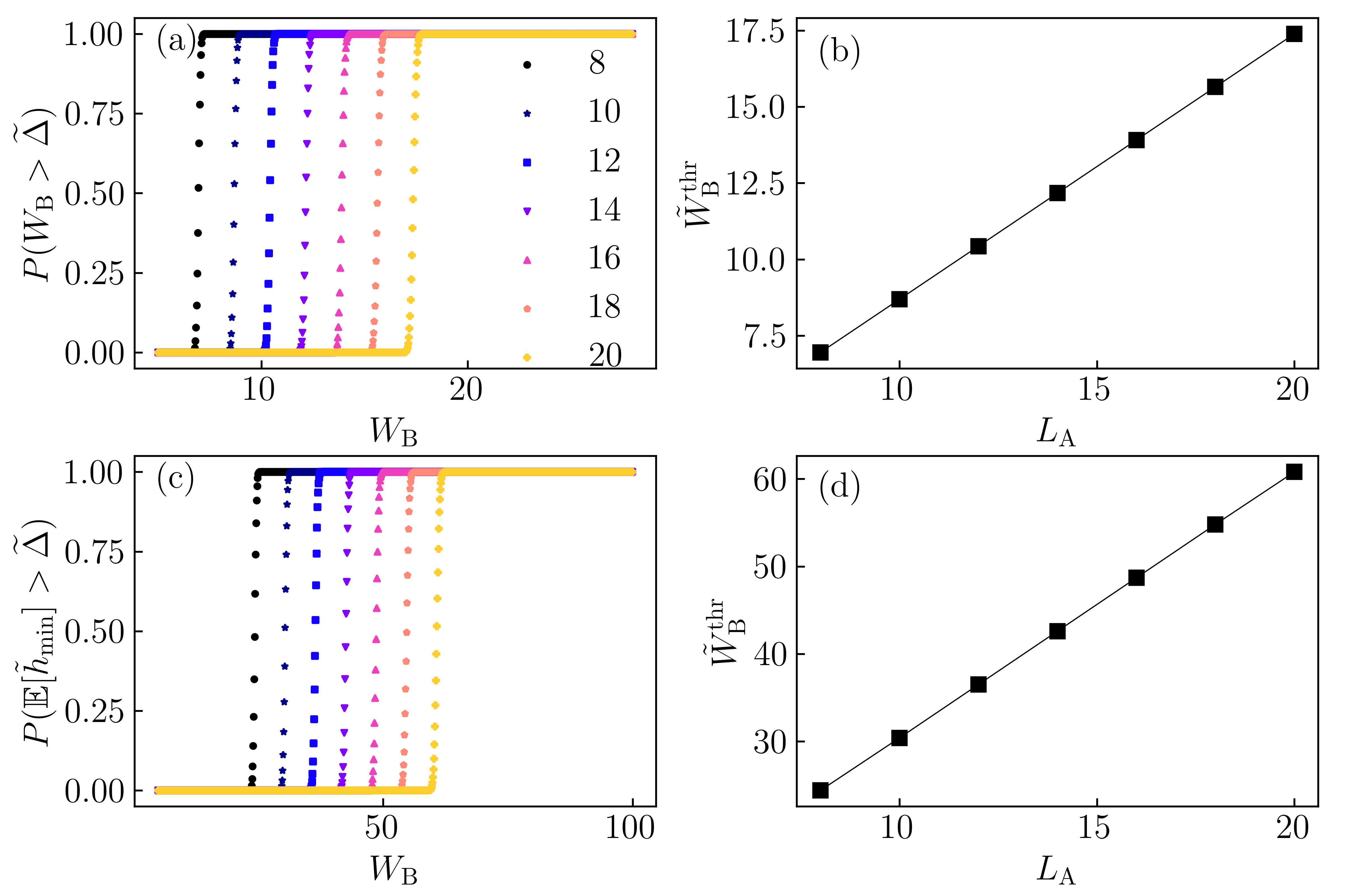}
        \caption{$L_\B = 1$: (a) $P(W_\B>\widetilde{\Delta})$ versus $W_\B$ for various $L_\A$'s. Panel (b) shows as a function of $L_\A$ the threshold $\tilde{W}_\B^\mathrm{thr}$ above which $P(W_\B>\widetilde{\Delta})> 0$ (here chosen to be $0.5$). $L_\B = 4$: (c) $P(\mathbb{E}[\tilde{h}_\mathrm{min}]>\tilde{\Delta})$ versus $W_\B$ for various $L_\A$'s. Panel (d) shows as a function of $L_\A$ the threshold $\tilde{W}_\B^\mathrm{thr}$ above which $P(\mathbb{E}[\tilde{h}_\mathrm{min}]>\tilde{\Delta})> 0$. For all panels, $J=1$, $W_\A=0.1$.}
        \label{fig:Irwin_Hall}
    \end{center}
\end{figure}

Although we have discussed vanishing of FGR-rate based on the behavior of the delta functions in Eq.~\eqref{eq:Single_impurity_FGR}, one wonders what role the matrix elements $|A_{\beta \alpha}|$ play. To this end, we evaluated numerically, for a given disorder realization $\{h_i^\A\}$ with given $L_\A$ and $L_\B=1$, the non-zero matrix elements $|A_{\beta \alpha}|^2$, $|A_{\beta \alpha}'|^2$ for allowed transitions between states $|\beta\rangle$ and $|\alpha\rangle$ and considered the distribution of the corresponding $\omega_{\beta \alpha}$'s. A systematic way to proceed is as follows: 
\begin{enumerate}
    \item Considering all possible eigenstates $|\alpha \rangle$, $| \beta \rangle \in Sp(\hat{H}_{L_\A})$ and with $\langle \alpha|\hat{S}^z|\alpha \rangle = \langle \beta|\hat{S}^z|\beta \rangle$, enumerate those eigenstates that yield $|A_{\beta \alpha}|^2 \ne 0$ and $|A_{\beta \alpha}'|^2 \ne 0$. In Fig.~\ref{fig:App_4}(a), we demonstrate that the eigenstates $|\alpha \rangle$ that belong to the representative $S^z = 0$ sector of the Hamiltonian $\hat{H}_{L_\A}$ are spread across the energy spectrum. For such eigenstates, compute $\omega_{\beta \alpha}$. One would end up with several possible $\omega_{\beta \alpha}$'s, yielding a probability distribution $P(\omega_{\beta \alpha})$. We find that the distribution $P(\omega_{\beta \alpha})$ shows a single peak and a finite spread around it, as shown in the main plot of Fig.~\ref{fig:App_4}(b) for $10$ different disorder realizations. 
    \item Use the distribution $P(\omega_{\beta \alpha})$ to estimate the condition for having inhibited spin transport, that is, for obtaining vanishing FGR-rate.
    \item To perform the aforementioned estimation, one may proceed as follows: Let us use $\tilde{\omega}_{\beta \alpha} \equiv \overline{\omega_{\beta \alpha}^2} - (\overline{\omega_{\beta \alpha}})^2$, with the overbar denoting numerical average, as a measure of the width of the distribution $P(\omega_{\beta \alpha})$. We calculate $\tilde{\omega}_{\beta \alpha}$ for $1000$ disorder realizations. The inset in Fig.~\ref{fig:App_4}(b) shows that the distribution of $\tilde{\omega}_{\beta \alpha}$ has a single peak and a finite spread around it. It is therefore evident that the non-self-averaging effects discussed for the quantity $\widetilde{m}_\B$ are not manifested here; in other words, the behavior of the quantity $\tilde{\omega}_{\beta \alpha}$ is typical across different disorder realizations, and it suffices for further analysis to consider its value for a typical disorder realization. Considering the behaviour of $\tilde{\omega}_{\beta \alpha}$ as a function of $L_\A$, we observe a monotonic growth with $L_\A$, as may be seen from Fig.~\ref{fig:App_4}(c). This growth suggests that for larger $L_\A$, one needs to tune $W_\B$ to higher values to have inhibited spin transport. Such a conclusion is in full agreement with the one arrived at from the analysis of the behaviour of delta functions.
\end{enumerate}

\subsection{For general $L_\B$}
The treatment until now applies to the case $L_\B=1$. We now argue that the conclusions arrived at also hold when one has $1<L_\B \ll L_\A$. The corresponding FGR-rate is (see Appendix~\ref{App-FGR-derivation})
\begin{align}\label{eq:FGR-general_La}
     &\Gamma_{\alpha \to [\beta]} \propto \sum_\beta \sum_{j=0}^{L_\B-2} \biggl[|A^{(j)}_{\beta\alpha}|^2 ~\delta^{(j)}_+ + |A'^{(j)}_{\beta \alpha}|^2 ~\delta^{(j)}_- \biggr] \nonumber \\
    &+ |A^{(L_\B')}_{\beta \alpha}|^2 \delta(\omega_{\beta \alpha} + h_{L_\B'}^\B) + |A'^{(L_\B')}_{\beta \alpha}|^2 \delta(\omega_{\beta \alpha} - h_{L_\B'}^\B),
\end{align}
where $\tilde{h}_j \equiv (h_j^\B - h_{j+1}^\B)$, $\delta^{(j)}_\pm \equiv \delta(\omega_{\beta \alpha} \pm \tilde{h}_j)$, $A^{(j)}_{\beta \alpha} = \langle \beta|\hat{S}_{j}^+\hat{S}_{j+1}^-|\alpha \rangle$, $A'^{(j)}_{\beta \alpha} = \langle \beta|\hat{S}_{j}^-\hat{S}_{j+1}^+|\alpha\rangle$ and $L_\B' \equiv L_\B-1$. Evidently, $\Gamma$ is zero provided $\tilde{h}_\mathrm{min} \equiv \min_j\{|\tilde{h}_j|\}>\Delta$, which arguing as before is equivalently the condition $\tilde{h}_\mathrm{min}>\tilde{\Delta}$; in view of $\tilde{h}_\mathrm{min}$ being a random variable, we may read the condition in terms of the average of $\tilde{h}_\mathrm{min}$ as $\mathbb{E}[\tilde{h}_\mathrm{min}]>\tilde{\Delta}$. In the following, using extreme value theory, we obtain the distribution $p(h_\mathrm{min})$ and consequently the desired expectation value $\mathbb{E}[\tilde{h}_\mathrm{min}]$. 

\subsubsection{Extreme value theory: derivation of $\mathbb{E}[\tilde{h}_\mathrm{min}]$ }
From the expression of the FGR rate in Eq.~\eqref{eq:FGR-general_La}, it is evident that all the delta functions will contribute zero under the condition $\tilde{h}_\mathrm{min}= \min_j \{\tilde{h}_j\} > \widetilde{\Delta}$, where we have $\tilde{h}_j = |h_j^\B - h_{j+1}^\B|$; $j = 0,\ldots, L_{\B} -2$.  Now, notice that $h_j^\B$s are themselves random variables drawn from a uniform distribution over the range $[-W_\B,W_\B]$. It follows straightforwardly that their absolute gaps, $\tilde{h}_j$, which are independent and identically-distributed (i.i.d.) variables just as $h_j^\B$'s, will follow the half-triangular distribution 
\begin{align}
    p(\tilde{h}_j)= \frac{1}{2W_\B^2}\begin{cases}
        2 W_\B - \tilde{h}_j;& ~ 0 \leq \tilde{h}_j \leq 2W_\B, \\
        0;& ~\mathrm{otherwise}.
    \end{cases}
\end{align}
To get the distribution of $\tilde{h}_\mathrm{min}$, we look at the following probability:
\begin{align}{\label{eq:h_min}}
    P(\tilde{h}_\mathrm{min} > h) = P(\tilde{h}_0 > h)~P(\tilde{h}_1 > h) \ldots P(\tilde{h}_{L_\B-2} > h), 
\end{align}
where the invoked factorization is justified since all the gaps are i.i.d. random variables. It is evident that each factor $P(\tilde{h}_j > h)$ can be expressed in terms of the cumulative distribution of $\tilde{h}_j$, as $P(\tilde{h}_j > h) = 1 - P(\tilde{h}_j \leq h) $. Since all factors $P(\tilde{h}_j > h)$ are identical, we obtain from Eq.~(\ref{eq:h_min}) that
\begin{align}
    P(\tilde{h}_\mathrm{min} > h) & = \left[1 - P(\tilde{h}_j \leq h)\right]^{L_\B -1} \nonumber \\
    & = \begin{cases}
         \left[1 - \frac{4W_\B h - h^2 }{4 W_\B^2}  \right]^{L_\B -1} ;~ & 0 \leq h \leq 2 W_\B, \\
         0;~ & \mathrm{otherwise}.
    \end{cases}
\end{align}

The probability distribution function (pdf) for $\tilde{h}_\mathrm{min}$ is obtained by taking the derivative of $P(\tilde{h}_\mathrm{min} \leq h)= 1 - P(\tilde{h}_\mathrm{min} > h)$, and it reads as 
\begin{align}
    p(\tilde{h}_\mathrm{min}) & =  \begin{cases}
        \frac{(L_\B-1)}{W_\B} f_-^{2L_\B -3} ; ~ & 0 \leq \tilde{h}_\mathrm{min} \leq 2W_\B, \label{appeq10} \\
        0 ~; ~ & \mathrm{otherwise},
    \end{cases}
\end{align} 
where we have defined $f_- \equiv (2W_\B - \tilde{h}_\mathrm{min})/(2 W_\B)$. 

The expectation value $\mathbb{E}[\tilde{h}_\mathrm{min}]$, which is required for $P(\mathbb{E}[\tilde{h}_\mathrm{min}]> \tilde{\Delta})$, is given by 
\begin{align} \label{eq:E_hmin}
    \mathbb{E}[\tilde{h}_\mathrm{min}] & = \int p(\tilde{h}_\mathrm{min})~   \tilde{h}_\mathrm{min}~\mathrm{d}\tilde{h}_\mathrm{min}\\
    & = \frac{2 W_\B}{2 L_\B -1}.
\end{align}
One may also obtain 
all the moments $\mathbb{E}[(\tilde{h}_\mathrm{min})^n]$ as
\begin{align}
    \mathbb{E}[(\tilde{h}_\mathrm{min})^n] = \frac{8 W_\B^n (L_\B -1) \Gamma(2L_\A -2) \Gamma(n+1)}{\Gamma(2L_\A + n -1)}.
\end{align}

Finally, coming back to  our earlier analysis, the condition for $\Gamma$ being zero is to have $P(\mathbb{E}[\tilde{h}_\mathrm{min}]>\tilde{\Delta})$, obtained by replacing $W_\B$ in Eq.~\eqref{eq:P-expression} with the expression of $\mathbb{E}[\tilde{h}_\mathrm{min}]$ in Eq.~\eqref{eq:E_hmin}.  Panels (c), (d) of Fig.~\ref{fig:Irwin_Hall} depict similar results for $L_\B = 4$ as for $L_\B=1$. We thus conclude that (i) regardless of the value of $L_\B$, our analytical treatment implies that the threshold $\tilde{W}_\B^\mathrm{thr}$ to observe inhibited spin-transport increases with increasing $L_\A$, and (ii) that the fact is in qualitative agreement with our earlier numerical finding, see Fig.~\ref{fig:fig_2}(j); Remarkably, the variation of the threshold is observed over similar range of $L_\A$-values in both theory and numerics.

\section{Outlook}\label{sec:conclusion}
In summary, we have established that thermalization in a disordered spin chain in contact with a weakly-disordered (or clean) chain is not always ``typical" and may crucially depend on disorder realizations and the size of the weakly-disordered chain. Moreover, for fixed $L_\A$ and $L_\B$, our analysis gives access to the regime of disorder strengths for which one can rely on single-sample measurement to infer on the occurrence of thermalization, or absence thereof. As mentioned earlier, our results pertain to finite-size systems, whereas in the true thermodynamic limit with fixed $L_A/L_B$ and $L_A, L_B \to \infty$, the non-self-averaging regime (II) is expected to disappear. Nevertheless, our findings remain highly significant, as virtually all experiments are conducted in finite-size settings. A notable example is the inhomogeneously disordered optical lattice implemented in Ref.~\cite{PhysRevX.9.041014} to investigate the avalanche picture. From this pragmatic standpoint, where both $L_A$ and $L_B$ are fixed, and far from the thermodynamic limit, our results identify a regime in which a single-sample measurement may be misleading due to strong sample-to-sample fluctuations.

In passing, we also remark that two clean Heisenberg chains with two different onsite fields and in contact become non-integrable, and exhibit Wigner-Dyson statistics for level spacing, leading to trivial thermalization~\cite{cipolloni_ETH_proof, Deutsch_ETH_original}. In stark contrast, if one of the chains (or both) is disordered, the level-spacing statistics crosses over from Wigner-Dyson to Poissonian, as demonstrated in Appendix~\ref{App-Level-spacing}. When Wigner-Dyson is not applicable and ETH might fail~\cite{MBL_Review_Nandkishore}, our work unveils how transport provides a probe to decide on thermalization.

Our work opens up several future directions:
\begin{itemize}
    \item While the observed growth of $W_\B^{\mathrm{thr}}$ with increasing $L_\A$ could be argued on physical grounds, the emergence of regime II is somewhat surprising. Our results imply that the crossover from absolute thermalization (regime I) to absolute localization (regime III) is not smooth but instead passes through an intermediate regime where neither thermalization nor localization is “typical.” We attribute this non-trivial behaviour to the full unitary evolution of the combined system. A promising direction for future investigation is to quantitatively assess how modeling part $\A$ as a GOE Hamiltonian, or approximating it as a Markovian bath—as commonly done in avalanche-based studies~\cite{David_luitz_PRL,Bath_Induced_Delocalization_Dries}—yields results that deviate from the ones obtained under exact unitary dynamics.

    \item While we have provided a perturbative analysis in Section~\ref{sec:analytics} to account for the inhibition of transport in the strong-disorder regime, a more refined microscopic understanding of spin transport across all disorder regimes remains an interesting avenue for further exploration. Additionally, studying our inhomogeneously-disordered model within a non-equilibrium setup, as in Ref.~\cite{rigol_transport}, could offer deeper insights into general transport phenomena.

    \item In our set-up, we have considered that the two chains are coupled via a single boundary spin. It would be interesting to explore if and how the qualitative picture conveyed in Fig.~\ref{fig:setup}(b) is modified when (i) the two chains are connected through multiple spins, or, (ii) chain $\B$ features long-range interactions among its constituents. In particular, it is worth investigating whether such modifications can enhance transport in the strong-disorder regime.

\end{itemize}

Finally, much like the experimental verification of the avalanche picture~\cite{leonard,PhysRevX.9.041014}, our results and questions can be readily investigated in ultracold atom platforms~\cite{Immanuel_Bloch_review_ultracold}, where Feshbach resonances~\cite{Petrov_feshbach} allow for precise control of interaction strengths.

\section{Acknowledgements} The authors thank Michael Kastner and Alessandro Campa for their valuable comments on the manuscript. S.G. acknowledges the financial support of the Department of Atomic Energy, Government of India, under Project Identification No. RTI 4002, and thanks ICTP-Abdus Salam International Centre for Theoretical Physics, Trieste, Italy, for support under its Regular Associateship scheme. S.K.P. is supported through a graduate fellowship at TIFR, funded by the Department of Atomic Energy (DAE), India.   S.K.P. acknowledges the ``Stability of quantum matter in and out of equilibrium at various scales" conference at the International Centre of Theoretical Sciences (ICTS), Bengaluru, India, where a part of this work was done. We gratefully acknowledge generous allocation of computing resources by the Department of Theoretical Physics (DTP) of the Tata Institute of Fundamental Research (TIFR), and technical assistance from Kapil Ghadiali and Ajay Salve. 

\section{Data Availability}
The data that support the findings of this study are
openly available at Zenodo at the link in Ref.~\cite{Data}.

\clearpage
\noindent

\appendix

\section{Equivalent of Fig. 1 for the case when systems A and B have disordered fields sampled from a Gaussian distribution instead of a uniform  distribution}\label{APP-Gaussian}

Here, we demonstrate that one arrives at the same conclusions as in the main text if the disorder distribution in Hamiltonian \eqref{eq:H} is taken to be Gaussian instead of the uniform distribution.

\begin{figure}[!h]
    \begin{center}
        \includegraphics[width = 8.6cm]{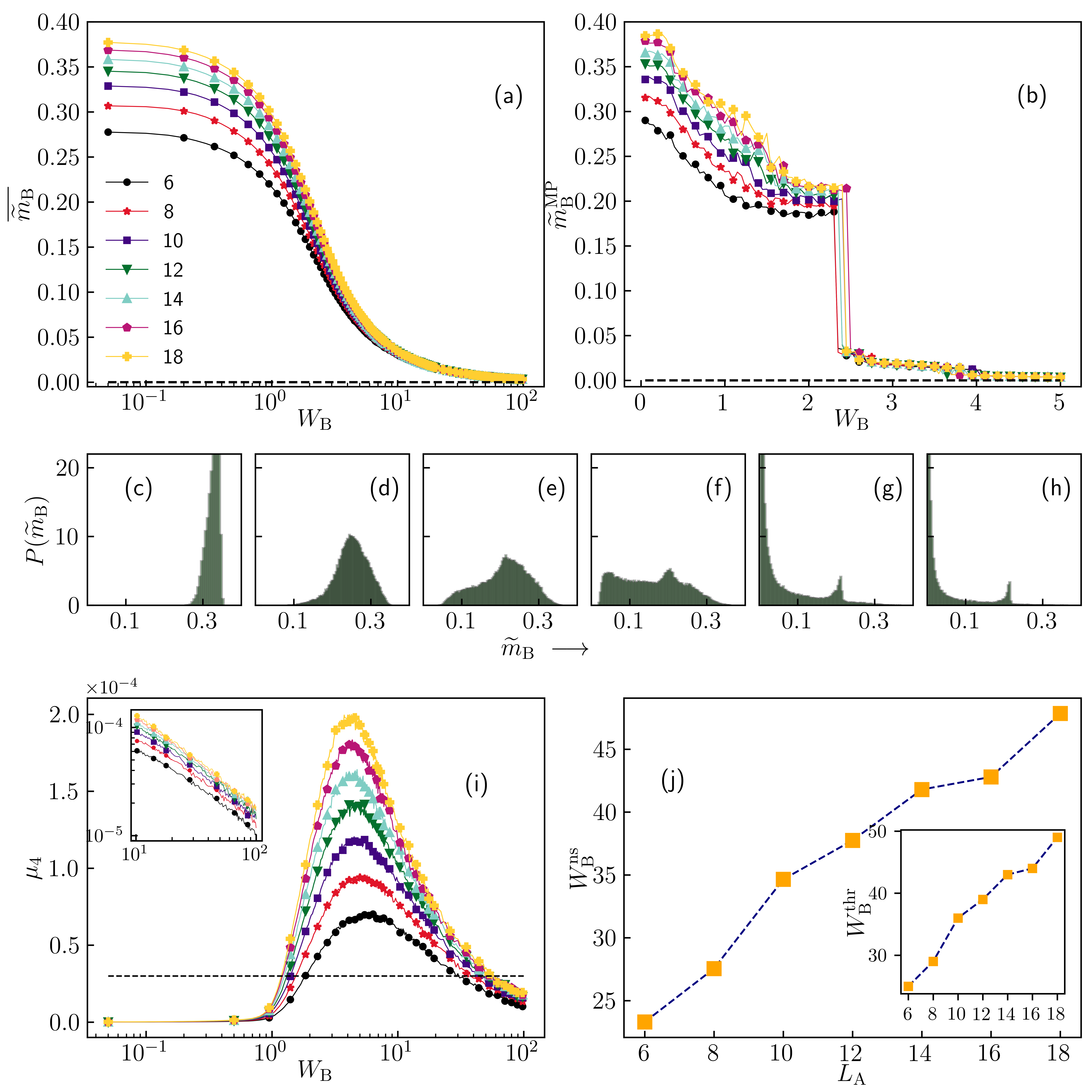} 
        \caption{The different panels in the figure show the same quantities as in Fig.~\ref{fig:fig_2} , but with $h_i^\A$, and $h_j^\B$ sampled from  Gaussian distributions $\mathcal{N}(0, W_\A)$, and $\mathcal{N}(0, W_\B)$, respectively. We fix $W_\A=0.1$. The figure shows that the information conveyed by Fig.~\ref{fig:fig_2} also holds when the choice of distribution for the fields $h_j^\B$ is changed from uniform to Gaussian.}
        \label{fig:App_1}
    \end{center}
\end{figure}

\section{Equivalent of Fig. 1 for the case when system A is clean with constant field $h^\A$} \label{App-uniform-field}
Here, we demonstrate that one arrives at the same conclusions as in the main text if the part $\A$ in Hamiltonian \eqref{eq:H} has a constant field instead of random fields sampled from a uniform distribution.
\begin{figure}[!htbp]
    \begin{center}
        \includegraphics[width = 8.6 cm]{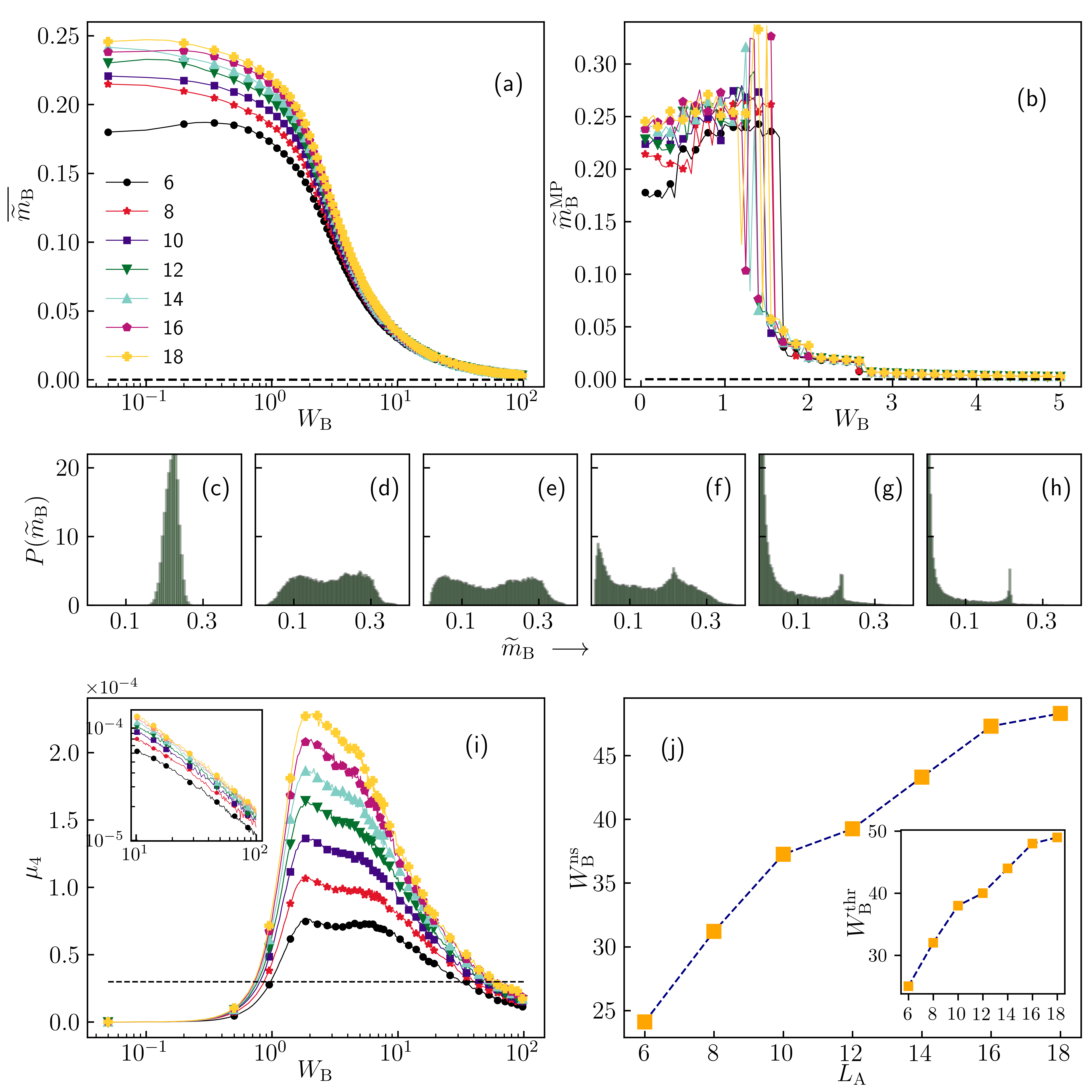}
        \caption{The different panels in the figure show the same quantities as in Fig.~\ref{fig:fig_2}, but with system $\A$ having constant onsite fields $h^\A=1$ instead of a disordered onsite field $h_i^\A$ as in Fig.~\ref{fig:fig_2} . The figure shows that the information conveyed by Fig.~\ref{fig:fig_2} also holds when system $\A$ has a constant onsite field.}
        \label{fig:App_2}
    \end{center}
\end{figure}

\section{Equivalent of Fig. 1 for the case when system A has uniform disorder but the system has periodic boundary conditions} \label{App-periodic-uniform-field}

\begin{figure}
    \begin{center}
        \includegraphics[width = 8.6 cm]{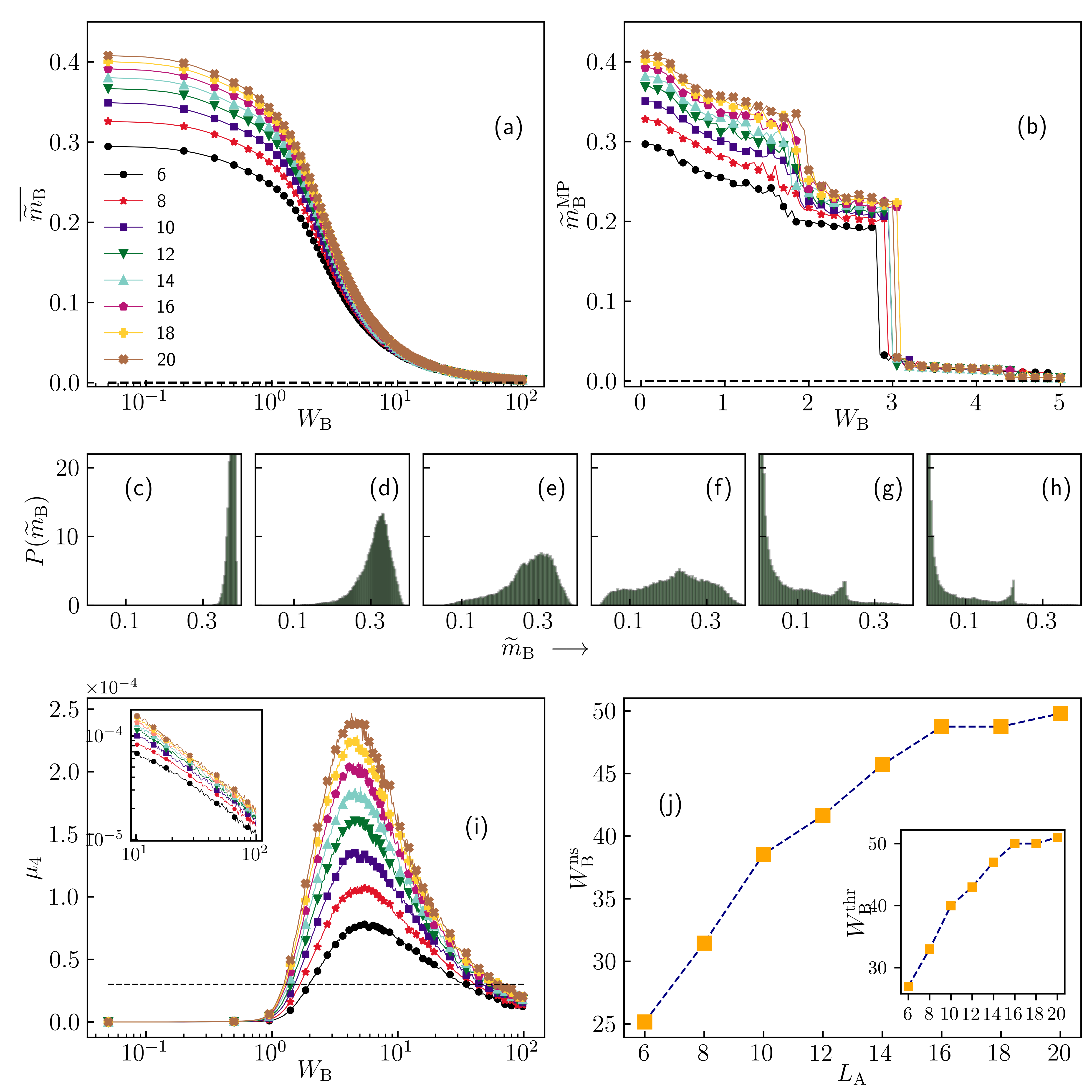}
        \caption{The different panels in the figure show the same quantities as in Fig.~\ref{fig:fig_2}, but with the system having periodic boundary conditions i.e, there are two points of contact between chains A and B. $h_i^\A$ is sampled from a uniform distribution of width $W_\mathrm{A} = 0.1$. The figure shows that the information conveyed by Fig.~\ref{fig:fig_2} also holds when the system has periodic boundaries instead of open boundaries.}
        \label{fig:App_3}
    \end{center}
\end{figure}

\section{Derivation of Eqs.~(\ref{eq:Single_impurity_FGR}) and~(\ref{eq:FGR-general_La}) } \label{App-FGR-derivation}
We start with the model Hamiltonian in Eq.~\eqref{eq:H_P},
\begin{align}
    \hat{H} = J \sum_{i=1}^{L_\A-1} \hat{\Vec{S_i}} \cdot \hat{\Vec{S}}_{i+1} + \sum_{i=1}^{L_\A} h_i^\A \hat{S_i^z}  + J \hat{\Vec{S_0}} \cdot \hat{\Vec{S}}_{1}+ W_\B \hat{S_0^z}.
\end{align}
Absorbing the $\hat{S}_0^z \hat{S}_1^z$-term into the summation and using the raising-lowering operators $\hat{S}_i^{\pm} = \hat{S}_i^x \pm i \hat{S}_i^y$, we can rewrite the Hamiltonian as
\begin{align}
    \hat{H} &= \underbrace{J \sum_{i=1}^{L_\A-1} \left( \hat{S}_i^x \hat{S}_{i+1}^x + \hat{S}_i^y \hat{S}_{i+1}^y \right) + J \sum_{i=0}^{L_\A-1} \hat{S}_i^z \hat{S}_{i+1}^z + \sum_{i=1}^{L_\A} h_i^\A \hat{S_i^z}}_{\hat{H}_{L_\A}} \\ \nonumber 
    &  + W_\B \hat{S}_0^z + \frac{J}{2} \left(\hat{S}_0^- \hat{S}_1^+ + \hat{S}_0^+ \hat{S}_1^- \right).
\end{align}
Invoking the interaction picture with respect to the term $W_\B \hat{S}_0^z$, the interaction Hamiltonian reads as $\hat{H}_{I} = e^{i W_\B \hat{S}_0^z t} (\hat{H} - W_\B \hat{S}_0^z) e^{-i W_\B \hat{S}_0^z t}$, which may be evaluated as follows. It is easily seen that one has $[\hat{H}_{L_\A}, \hat{S}_0^z] = 0$. Thus, the only non-trivial term to be evaluated is $e^{i W_\B \hat{S}_0^z t} \frac{J}{2} \left(\hat{S}_0^- \hat{S}_1^+ + \hat{S}_0^+ \hat{S}_1^- \right) e^{-i W_\B \hat{S}_0^z t}$, which can be done using the well-known Baker-Campbell-Hausdorff identity:
\begin{align}
    e^{i \lambda \hat{X}} \hat{A} e^{-i \lambda \hat{X}} = \hat{A} + (i \lambda)[\hat{X}, \hat{A}] + \frac{(i \lambda)^2}{2!} [\hat{X}, [\hat{X}, \hat{A}]] + \dots;
\end{align}
We get
\begin{align}
    \hat{H}_{I} = \hat{H}_{L_\A} + \underbrace{ \frac{J}{2} \biggl(e^{-i W_\B t} \hat{S}_1^+ \hat{S}_0^- + e^{i W_\B t} \hat{S}_1^- \hat{S}_0^+ \biggr)}_{\hat{H}^\prime(t)}.
\end{align}

To proceed, we treat $\hat{H}^\prime(t)$ as a weak perturbation to the Hamiltonian $\hat{H}_{L_\A}$ in calculating the Fermi's Golden Rule (FGR) transition rate between the eigenstates of $\hat{H}_{L_\A}$, specifically, for a transition from eigenstate $| \alpha \rangle$ to eigenstate $| \beta \rangle$. One may proceed as follows: one prepares the system in the state $|\alpha\rangle$ at time $t_0=-\infty$, and then turns on the time-dependent perturbation $\hat{H}^\prime(t)$ at time $t_0$, which in course of time induces transitions between the various eigenstates of $\hat{H}_{L_\A}$. In the first-order perturbation theory, the corresponding transition amplitude is given by (see any standard reference on quantum mechanics, e.g., Ref.~\cite{sakurai})
\begin{align}
    c_{\beta \alpha}^{(1)} (t) &= \frac{-i J}{2}  ~\lim_{t_0 \to -\infty} \int_{t_0}^{t} \langle \beta| e^{i \hat{H}_{L_\A}t'}  \hat{H}^\prime(t') e^{-i \hat{H}_{L_\A}t'}|\alpha \rangle \mathrm{d}t'.
\end{align}    
Instead of turning on the perturbation instantly at $t_0$, we may for convenience of evaluation of $c_{\beta \alpha}^{(1)} (t)$ turn it on slowly in time, which is implemented by introducing the factor $e^{\eta t}$, as 
\begin{align}
    c_{\beta \alpha}^{(1)} (t) &= \frac{-i J}{2} \lim_{\substack{\eta \to 0 \\ t_0 \to -\infty}}
\int_{t_0}^{t} \langle \beta| e^{i \hat{H}_{L_\A}t'}  \hat{H}^\prime(t') e^{-i \hat{H}_{L_\A}t'}|\alpha \rangle e^{\eta t'} \mathrm{d}t' \nonumber \\
                     &= \frac{-i J}{2}  ~\lim_{\substack{\eta \to 0 \\ t_0 \to -\infty}}
 \biggl({A_{\beta \alpha}} e^{i \omega_{\beta \alpha}t' - i W_\B t' + \eta t'} \nonumber \\  & 
                     + {A'_{\beta \alpha}} e^{i \omega_{\beta \alpha}t' + i W_\B t'+ \eta t'} \biggr) \mathrm{d}t' \nonumber \\ 
                     &= \frac{-i J}{2} \lim_{\eta \to 0} \biggl[A_{\beta \alpha} \frac{e^{i \omega_{\beta \alpha}t - i W_\B t + \eta t}}{(i \omega_{\beta \alpha} - i W_\B + \eta)} \\ \nonumber &  +  A'_{\beta \alpha} \frac{e^{i \omega_{\beta \alpha}t + i W_\B t+ \eta t}}{(i \omega_{\beta \alpha} + i W_\B + \eta)}\biggr],
\end{align}
where $ A_{\beta \alpha} \equiv \langle \beta| \hat{S}_1^+ \hat{S}_0^-| \alpha \rangle$, and $A'_{\beta \alpha} = \langle \beta| \hat{S}_1^- \hat{S}_0^+|\alpha \rangle$. 
This gives us 
\begin{align}
    |c_{\beta \alpha}^{(1)}(t)|^2 &= \frac{J^2}{4} \lim_{\eta \to 0} e^{2 \eta t} \biggl[|A_{\beta \alpha}|^2 \frac{1}{\eta^2 + (\omega_{\beta \alpha} - W_\B)^2} \\ \nonumber & + |A'_{\beta \alpha}|^2 \frac{1}{\eta^2 + (\omega_{\beta \alpha} + W_\B)^2} \biggr].
\end{align}
The total FGR transition rate out of state $|\alpha \rangle$, given by $\Gamma_{\alpha \to [\beta]} = \sum_{\beta \neq \alpha} \frac{\mathrm{d}}{\mathrm{d}t} |c_{\beta \alpha}(t)|^2$, is then obtained in the limit $t \to \infty$ as
\begin{align}
    \Gamma_{\alpha \to [\beta]} &\propto \sum_\beta \Biggl \{|A_{\beta \alpha}|^2 \delta(\omega_{\beta \alpha} +W_\B) + |A'_{\beta \alpha}|^2 \delta(\omega_{\beta \alpha} -W_\B ) 
                        \Biggr \},
\end{align}
which is Eq.~\eqref{eq:Single_impurity_FGR} of the main text.

In order to derive Eq.~\eqref{eq:FGR-general_La} of the main text, let us consider the case when the system B has more than one site, such that $L_\B$, the number of sites in system $\B$ satisfies $1<L_\B \ll L_\A$. In this case also there is a timescale separation between the dynamics governed by the Hamiltonians $\hat{H}^\prime$ and $\hat{H}_{L_\A}$:  $t_{\hat{H}^\prime}/t_{\hat{H}_{L_\A}} = ||\hat{H}_{L_\A}||/||H^\prime|| \gg 1$. We label the spins in system B as: The rightmost spin is labelled as $0$, the one to its left as $1$ and so on up to $L_\B - 1$. The spin labelled as $L_\B$ in the expressions below is the first spin in system A, and it is this spin  that is shared between systems A and B. Now, the system Hamiltonian will read as
\begin{align}
    \hat{H} &= \hat{H}_{L_\A} + \frac{J}{2} (\hat{S}_{L_\B-1}^+ \hat{S}_{L_\B}^- + \hat{S}_{L_\B-1}^- \hat{S}_{L_\B}^+) \nonumber\\ & 
    + \frac{J}{2} \sum_{i=0}^{L_\B-2} ( \hat{S}_i^+ \hat{S}_{i+1}^- + \hat{S}_i^- \hat{S}_{i+1}^+) + \sum_{i=0}^{L_\B-1} h^\B_i \hat{S}_i^z.
\end{align}
In the interaction picture with respect to $\sum_{i=0}^{L_B-1} h_i^\B \hat{S}_i^z$, the interaction Hamiltonian reads as
\begin{align}
    \hat{H}_I &= \hat{H}_{L_\A} + \frac{J}{2} (e^{i h^\B_{L_\B-1} t} \hat{S}_{L_\B-1}^+ \hat{S}_{L_\B}^- + e^{-i h^\B_{L_\B-1}t}\hat{S}_{L_\B-1}^- \hat{S}_{L_\B}^+) \nonumber \\ & 
    + \frac{J}{2} \sum_{i=0}^{L_\B-2} \biggl( e^{i (h^\B_i - h^\B_{i+1}) t} \hat{S}_i^+ \hat{S}_{i+1}^- + e^{i (-h^\B_i + h^\B_{i+1}) t}\hat{S}_i^- \hat{S}_{i+1}^+ \biggr).
\end{align}
It is then clear that following the same procedure as followed above for deriving Eq.~\eqref{eq:Single_impurity_FGR} , we will get the FGR
rate to be of the form
\begin{align}
    \Gamma_{\alpha \to [\beta]} & \propto \sum_\beta \sum_{j=0}^{L_\B-2} \biggl[|A_{\beta \alpha}^{(j)}|^2 ~\delta(\omega_{\beta \alpha}  + (h^\B_j - h^\B_{j+1})) \\ \nonumber 
    &+ |A'^{(j)}_{\beta \alpha}|^2 ~\delta(\omega_{\beta \alpha} - (h^\B_j - h^\B_{j+1}))\biggr]  \\ \nonumber 
    &+ |A_{\beta \alpha}^{(L_\B-1)}|^2 \delta(\omega_{\beta \alpha} + h^\B_{L_\B-1}) \nonumber \\ & 
    + |A'^{(L_\B-1)}_{\beta \alpha}|^2 \delta(\omega_{\beta \alpha} - h^\B_{L_\B-1}),
\end{align}
where we have $A_{\beta \alpha}^{(j)} = \langle \beta|\hat{S}_{j}^+\hat{S}_{j+1}^-|\alpha \rangle$ and $A'^{(j)}_{\beta \alpha} = \langle \beta|\hat{S}_{j}^-\hat{S}_{j+1}^+|\alpha \rangle$. The above is Eq.~\eqref{eq:FGR-general_La} in the text.

\section{Generalized Irwin-Hall Distribution and derivation of Eq.~\eqref{eq:P-expression} } \label{APP-Irwin-Hall}
Consider $L$ random variables $X_i$ sampled independently from a uniform distribution $\in [a, b]$, with $a \geq 0$, $b >a$. Next, consider the sum $Z = \sum_{i=1}^L X_i$, and our aim is to calculate its cumulative distribution function (CDF). The moment generating function (MGF) of $Z$ is given by $M_Z(s) = \mathbb{E}[e^{sZ}] = \mathbb{E}[e^{s \sum_{i=1}^{L} X_i}] = \prod_{i=1}^{L} \mathbb{E}[e^{sX_i}] = \left( \mathbb{E}[e^{sX}] \right)^L = [M_X(s)]^L$, where we have used the fact that the $X_i$'s are independent and identically-distributed (i.i.d.) random variables. Let us make note of the fact that the MGF of the random variable $X$ and the Laplace transform of its probability density function (PDF) $p(x)=p_X(x)$ are related as:
\begin{align}
    \mathcal{L} \{p(x)\} = \int_0^{\infty} e^{-sx} p(x) ~\mathrm{d}x = M_X(-s),
\end{align}
since one has $p(x) = 0$ for $x < a$. We also note that $p(x)$ is the derivative of the CDF $F_X(x)$:
\begin{align}
    p(x) = \frac{\mathrm{d} F_X(x)}{\mathrm{d}x},
\end{align}
by virtue of the fact that $F_X(x) = \int_{0}^{x} p(t) ~\mathrm{d}t$. It then follows that the MFG and the CDF are related by
\begin{align}
   M_X(-s) =  \mathcal{L} \biggl \{ \frac{\mathrm{d} F(x)}{\mathrm{d}x} \biggr \} = s \mathcal{L} \{ F (x) \} - F(0),
\end{align}
which gives us the expression for the CDF as
\begin{align}
    F_X(x) = \mathcal{L}^{-1} \biggl \{ \frac{1}{s} M_X(-s) \biggr \},
\end{align}
since $F(0) = 0$. Using this to calculate the CDF of $Z$, we get:
\begin{align}
    F_Z(z) = \mathcal{L}^{-1} \biggl \{ \frac{1}{s} M_Z (-s) \biggr \} = \mathcal{L}^{-1} \biggl \{ \frac{1}{s} [M_X (-s)]^L \biggr \}.
\end{align}

Now, substituting for $M_Z(s)$ in the last equation, we get
\begin{align}
    F_Z(z) = \mathcal{L}^{-1} \biggl \{ \frac{1}{s} \biggl[ \frac{e^{-sb} - e^{-sa}}{-s(b-a)} \biggr]^L \biggr \} \qquad \mathrm{for} ~s \neq 0.
\end{align}
Taking the inverse Laplace transform yields the expression
\begin{align}
    F_Z(z) = \mathcal{L}^{-1} \biggl \{ \frac{1}{(b-a)^n s^{L+1}} \biggr[ \sum_{k=0}^{L} {L \choose k} (-1)^k e^{-s[(b-a)k + aL]}\biggr] \biggr\}.
\end{align}
Now, using the fact that $\mathcal{L}^{-1} \{ e^{-as} \mathcal{L} \{f \}\} = u_a(t) f(t-a)$, where $u_a(t)$ is the unit step 
 function, we get
\begin{align}
    F_Z(z) = \frac{1}{(b-a)^L ~L!} \sum_{k=0}^{L} {L \choose k} (-1)^k \biggl[z - (b-a)k - aL \biggr]_+,
    \label{eq:CDF_Irwin_hall}
\end{align}
where we have 
\begin{align}
    (x-\xi)_+ = \begin{cases}
                (x - \xi) & \mathrm{if} ~(x-\xi) \geq 0, \\
                \qquad 0 & \mathrm{if} ~(x-\xi) < 0.
                \end{cases}
\end{align}
Note that Eq.~\eqref{eq:CDF_Irwin_hall} holds for $z$ in the range $aL \leq z \leq bL$. From Eq.~\eqref{eq:CDF_Irwin_hall}, one obtains Eq.~\eqref{eq:P-expression}. The CDF for the full range of $z$ will be:
\begin{align}
            P_Z(z) = \begin{cases}
                0 & \mathrm{if} ~z < aL, \\
                F_Z(z) & \mathrm{if} ~aL \leq z \leq bL,\\
                1 & \mathrm{if} ~z > bL.
                \end{cases}
\end{align}
\section{Level-spacing statistics of two clean Heisenberg chains in contact}\label{App-Level-spacing}
\begin{figure}[!h]
    \begin{center}
        \includegraphics[width = 0.5\textwidth]{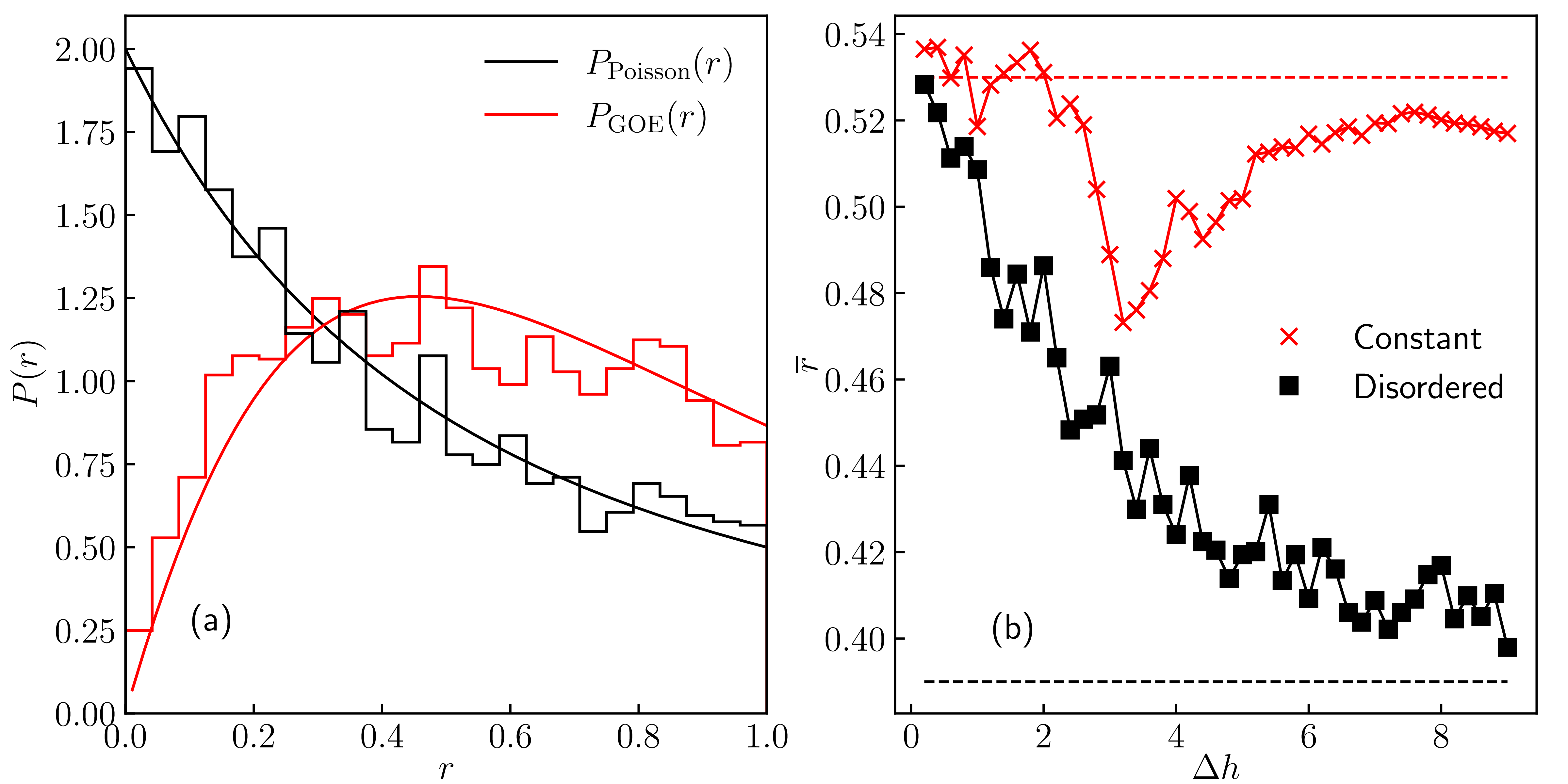}
        \caption{For $L_\B = 4$ and $L_\A = 10$, panel (a) shows a comparison of the distributions of $r$ for the case when sector B has constant onsite fields ($h^\B = 10.0$) and disordered onsite field ($W_\B = 10.0$). In the former case, the distribution fits to the distribution $P_{\mathrm{GOE}}(r)$, see Eq.~\eqref{eq:app-WD}, while in the latter case, the distribution fits to distribution $P_{\mathrm{Poisson}} (r)$, Eq.~\eqref{eq:app-Poisson}. Panel (b) shows how the average $\overline{r}$ crosses over from its value for Wigner-Dyson statistics to its value for Poissonian statistics, i.e., from the value of $0.53$ to the value of $0.39$,  with increasing $\Delta h = W_\B - h^\A$, for the disordered case (the black curve). On the other hand, for the case of constant onsite field (the red curve), the average $\overline{r}$ remains close to the value $0.53$ with increase in $\Delta h = h^\B - h^\A$.}
        \label{fig:App_5}
    \end{center}
\end{figure}

A quantum system is said to be chaotic if the normalized distribution of spacing between consecutive energy levels follows the Wigner-Dyson distribution \cite{BGS_conjecture,ETH_review}, an implication of which is that the Hamiltonian spectrum shows level repulsion. On the other hand, according to the Berry-Tabor conjecture \cite{berry_tabor}, level spacing for integrable systems follows a Poissonian distribution. Any non-integrable chaotic system is known to satisfy ETH and will thus thermalize~\cite{ETH_review,ETH_Srednicki}, meaning that 
its local observables will relax to microcanonical-equilibrium values. Such a scenario does not hold for integrable systems that do not follow ETH.\\
\indent Let us consider the case of two Heisenberg chains A and B with equal onsite fields $h^\A$ and $h^\B$, respectively, which are put in contact through a boundary spin shared between the two chains. The Hamiltonian of the full A+B system reads as
\begin{align}
    \hat{H} = J\sum_{i =1}^{L-1} \hat{\Vec{S_i}} \cdot \hat{\Vec{S}}_{i+1}  + h^\A \sum_{i \in \A}
    \hat{S_i^z} + h^\B \sum_{j \in \B}  \hat{S_j^z}
    \label{eq:H_constant}.
\end{align} 
We obtain the level statistics corresponding to the energy spectrum $\{E_\alpha\}$ of the Hamiltonian $\hat{H}$ in the $\hat{S}_z = 0$ sector. Confining to the states in the middle of the spectrum, we look at the statistics of the quantity $r_\alpha \equiv \min\{\delta_\alpha, \delta_{\alpha+1}\}/\max\{\delta_\alpha, \delta_{\alpha+1}\}$, with $\delta_\alpha \equiv E_\alpha - E_{\alpha+1}$. As seen in Fig.~\ref{fig:App_5}(a) for the case $h^\A=1.0$ and $h^\B=10.0$, we find the corresponding distribution $P(r)$ to be in excellent agreement with the Wigner-Dyson distribution $P_{\mathrm{GOE}}(r)$ corresponding to a Gaussian Orthogonal Ensemble (GOE), given by~\cite{Atas_level_spacing_ratio}
\begin{align}
    P_{\mathrm{GOE}}(r) = \frac{27}{4} \frac{r + r^2}{(1 + r + r^2)^\frac{5}{2}} \Theta(1-r),
    \label{eq:app-WD}
\end{align}
which yields the average value $\overline{r}= 0.5359$. In contrast, when system B has disordered onsite fields (field values for individual sites chosen independently and uniformly in $[-W_\B,W_\B]$, as considered in the main text, see Eq.~(1)), the level statistics of the entire system does not always satisfy the Wigner-Dyson distribution~\eqref{eq:app-WD}; instead, for strong disorder, the level statistics satisfies a Poissonian distribution $P_{\mathrm{Poisson}}(r)$~\cite{Atas_level_spacing_ratio}, see Fig.~\ref{fig:App_5}(a):
\begin{align}
    P_{\mathrm{Poisson}} (r) = \frac{2}{1 + r^2} \Theta(1-r),
    \label{eq:app-Poisson}
\end{align}
yielding the average $\overline{r}= 0.386$. As shown in Fig.~\ref{fig:App_5}(b), for this disordered case, the quantity $\overline{r}$ as a function of $\Delta h = (W_\B -h^\A)$ crosses over from the value of 0.535 to the value of 0.386 with increasing $\Delta h$. This implies that keeping the field values in system A fixed, as one makes system B more and more disordered, the entire system crosses over from being chaotic to integrable. In such a scenario, ETH is not guaranteed~\cite{MBL_Review_Nandkishore}, which calls for a detailed study devoted to unraveling of thermalization in the system at hand, which is what we report on in the main text. In passing, we show in Fig.~\ref{fig:App_5} that in contrast to the aforementioned disordered case, one has in the case of equal onsite fields in both systems A and B that the average $\overline{r}$ as a function of $\Delta h = (h^\B -h^\A)$ remains close to the value consistent with Wigner-Dyson distribution~\eqref{eq:app-WD} and does not show any trend of crossing over to the value consistent with the Poisson distribution~\eqref{eq:app-Poisson}. Thus, in this clean case, the entire system remains chaotic for all values of $\Delta h$, and hence, the $z$-magnetization per spin of system B is guaranteed to relax to equilibrium.  
\section{Comparison of the sixth, fourth and second central moments}\label{App-Compare_Var_Mu}

\begin{figure}[!h]
    \begin{center}
        \includegraphics[width = 0.5\textwidth]{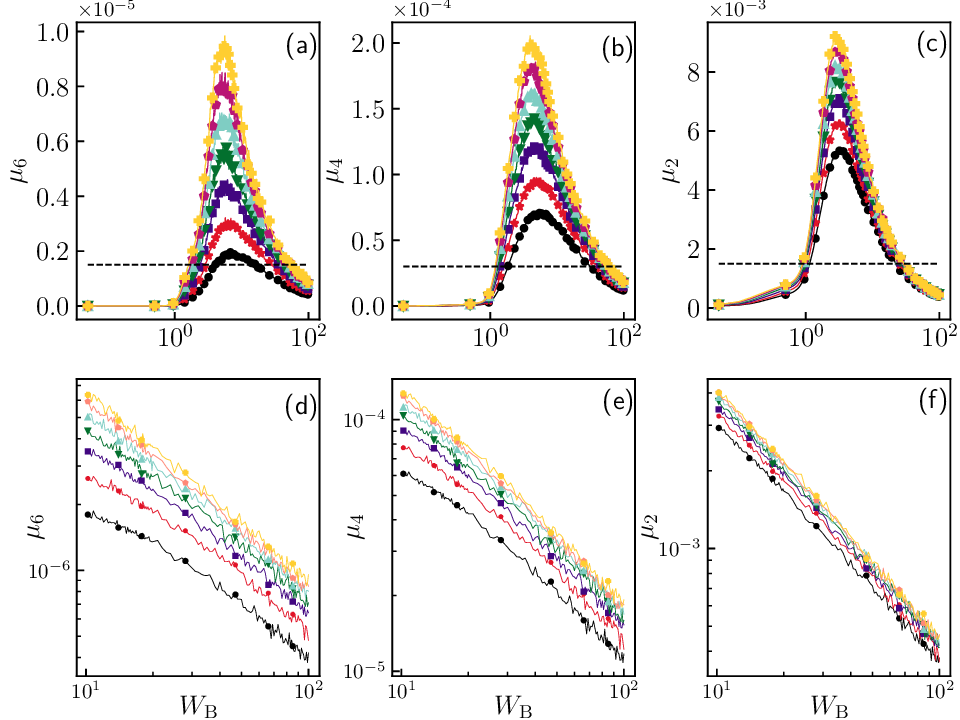}
        \caption{For the case when $h_i^\mathrm{A}$ and $h_i^\mathrm{B}$ are drawn from Uniform distributions $[-W_\A, W_\A]$ and $[-W_\B, W_\B]$ and with $W_\A = 0.1$, panels (a), (b) and (c) show the variation of $\mu_6$, $\mu_4$ and $\mu_2$ throughout the whole range of $W_\mathrm{B}$. Panels (d), (e) and (f) show a zoomed in image in the region where $W_\mathrm{B}$ is between $10$ and $10^2$.}
        \label{fig:App_6}
    \end{center}
\end{figure}

Identification of the threshold value, $W^\mathrm{thr}_\mathrm{B}$, requires the quantity of choice to be able to resolve well the behavior for different system sizes near the region where the quantity crosses the non-zero threshold (black dashed line in Fig. \ref{fig:fig_2} (i) and Fig. \ref{fig:App_6} (a), (b) and (c)). Figure~\ref{fig:App_6} displays a side-by-side comparison of $\mu_6\equiv \overline{(\widetilde{{m}}_\B-\overline{\widetilde{{m}}}_\B)^6}$, $\mu_4\equiv \overline{(\widetilde{{m}}_\B-\overline{\widetilde{{m}}}_\B)^4}$ and  $\mu_2\equiv \overline{(\widetilde{{m}}_\B-\overline{\widetilde{{m}}}_\B)^2}$, which clearly shows better resolution between different system sizes in the case of the higher moment. This helps in better identification of $W^\mathrm{thr}_\mathrm{B}$.

\newpage
\bibliographystyle{unsrturl}
\bibliography{main.bib}
\end{document}